\documentclass[12pt,a4paper,oneside,pdftex]{article}

\usepackage[english]{babel}
\usepackage[utf8]{inputenc}
\usepackage[T1]{fontenc}

\usepackage{lmodern}

\usepackage{latexsym}

\usepackage{geometry}

\newtheorem{EXAMPLE}{Example}[section]
\newtheorem{theorem}{Theorem}[section]

\newtheorem{lemma}[theorem]{Lemma}
\newtheorem{proposition}[theorem]{Proposition}
\newtheorem{corollary}[theorem]{Corollary}
\newtheorem{REMREM}{Remark}[section]
\newtheorem{DEFINITION}{Definition}[section]

\newenvironment{example}{\begin{EXAMPLE}\small}{\end{EXAMPLE}}

\newenvironment{definition}{\begin{DEFINITION}\rm}{\end{DEFINITION}}

\newcommand{\VAR}{\mbox{$Var$}}
\newcommand{\TRUE}{\mbox{$True$}}
\newcommand{\FALSE}{\mbox{$False$}}
\newcommand{\FUNC}[1]{\mbox{$Func_{#1}$}}
\newcommand{\PRED}[1]{\mbox{$Pred_{#1}$}}
\newcommand{\TERM}[1]{\mbox{$Term_{#1}$}}

\newcommand{\FEA}[1]{\mbox{$EQ_{#1}$}}

\newcommand{\EQNA}[1]{\mbox{$Eqn_{#1}$}}
\newcommand{\EQNB}[1]{\mbox{$Eqn_{#1}/_{\approx}$}}

\newcommand{\SUBEMPTYA}{\mbox{$\varepsilon$}}
\newcommand{\SUBEMPTYB}{\mbox{$\varepsilon_{\approx}$}}
\newcommand{\SUBFALSE}{\mbox{$\bot$}}
\newcommand{\SUBA}[1]{\mbox{$Sub_{#1}$}}
\newcommand{\SUBB}[1]{\mbox{$Sub_{#1}^{\bot}$}}
\newcommand{\SUBC}[1]{\mbox{$Sub_{#1}^{\bot}/{\approx}$}}

\newcommand{\TRUTH}[3]{\mbox{\( #1 \models_{#2} #3 \)}}
\newcommand{\NOTTRUTH}[3]{\mbox{\( #1 \not \models_{#2} #3 \)}}

\newcommand{\REWRA}[3]{\mbox{\( #1 \stackrel{#2}{\longrightarrow} #3 \)}}
\newcommand{\REWRB}[2]{\mbox{\( #1 \longrightarrow^{\star} #2 \)}}

\newcommand{\SOLN}[2]{\mbox{$soln_{#1}(#2)$}}
\newcommand{\INST}[2]{\mbox{$inst_{#1}(#2)$}}
\newcommand{\ANS}[3]{\mbox{$\langle #1 \rangle_{#2}^{#3}$}}

\newcommand{\BB}[2]{\mbox{$B_{#1}^{#2}$}}
\newcommand{\HA}{\mbox{$H$}}
\newcommand{\HB}{H}
\newcommand{\UU}[2]{\mbox{$U_{#1}^{#2}$}}
\newcommand{\VV}[2]{\mbox{$V_{#1}^{#2}$}}

\newcommand{\DOM}[1]{\mbox{$dom(#1)$}}
\newcommand{\RANGE}[1]{\mbox{$range(#1)$}}
\newcommand{\VARS}[1]{\mbox{$vars(#1)$}}
\newcommand{\ELIM}[1]{\mbox{$elim(#1)$}}
\newcommand{\PARAM}[1]{\mbox{$param(#1)$}}
\newcommand{\BOUND}[1]{\mbox{$bound(#1)$}}
\newcommand{\KERNEL}[1]{\mbox{$kernel(#1)$}}
\newcommand{\RESTR}[2]{\mbox{$#1 | #2$}}

\newcommand{\EQ}[1]{\mbox{${\cal E}$}-#1}
\newcommand{\BINDA}[2]{\mbox{$#1 \leftarrow #2$}}
\newcommand{\BINDB}[2]{\mbox{$\{ #1 \leftarrow #2 \}$}}
\newcommand{\DIFF}[2]{\mbox{$Diff(#1,#2)$}}
\newcommand{\CARD}[1]{\mbox{$|#1|$}}

\newcommand{\END}{\mbox{$\Box$}}
\newcommand{\IM}{\mbox{$\Phi$}}
\newcommand{\CCA}[3]{\mbox{$#1[#2 \leftarrow #3]$}}
\newcommand{\CCB}[3]{#1[#2 \leftarrow #3]}
\newcommand{\KK}[1]{\mbox{${\cal K}_{#1}$}}

\newcommand{\LA}[1]{\mbox{$L_{#1}$}}
\newcommand{\LB}[1]{L_{#1}}

\newcommand{\OLA}[2]{\mbox{$\overline{#1}#2$}}
\newcommand{\OLB}[2]{\overline{#1}#2}

\newcommand{\MANS}[3]{\mbox{$[#1]_{#2}^{#3}$}}
\newcommand{\EMPTYSEQA}{\mbox{$\Lambda$}}
\newcommand{\EMPTYSEQB}{\Lambda}

\newcommand{\PROJ}[2]{\mbox{$#1 | #2$}}

\usepackage[pdftex,bookmarksnumbered,pagebackref]{hyperref} 

\hypersetup{
 pdftitle={Conjunctive Queries, Existentially Quantified Systems of Equations and Finite Substitutions},
 pdfauthor={Ján Komara},
}

\hypersetup{
 unicode,
 colorlinks=true,
 allcolors=[rgb]{0,0.5,0.5},
 urlcolor=[rgb]{0.58,0.0,0.83},
 breaklinks=true,
}

\begin{document}

\title{
 Conjunctive Queries, Existentially Quantified Systems of Equations and Finite Substitutions%
 \thanks{Reprint of the technical report TR mff-ii-10-1992, September 1992.}
}

\author{
 {\Large Komara J\'{a}n}
 \and
 {\small Institute of Informatics, MFF UK, Mlynsk\'{a} dolina, 842 15 Bratislava, Czechoslovakia}
}

\date{}

\maketitle

\begin{abstract}
This report presents an elementary theory of unification for positive
conjunctive queries. A positive conjunctive query is a formula constructed from
propositional constants, equations and atoms using the conjunction $\wedge$ and
the existential quantifier $\exists$. In particular, empty queries correspond
to existentially quantified systems of equations --- called \EQ{formulas}. We
provide an algorithm which transforms any conjunctive query into a solved form.
We prove some lattice-theoretic properties of queries. In particular, the
quotient set of \EQ{formulas} under an equivalence relation forms a complete
lattice. Then we present another lattice --- a lattice of finite substitutions.
We prove that the both lattices are isomorphic. Finally, we introduce the
notion of application of substitutions to formulas and clarify its relationship
to \EQ{formulas}. This theory can be regarded as a basis for alternative
presentation of logic programming.
\end{abstract}

\section{Introduction}\label{sc:intr}
In this paper we present an elementary theory of unification for positive
conjunctive queries. A positive conjunctive query (or just a query for short)
is a natural generalization of a system of equations. It is a formula
constructed from propositional constants, equations and atoms using the
conjunction $\wedge$ and the existential quantifier $\exists$. In particular,
empty query corresponds to existentially quantified system of equations ---
called \EQ{formulas}. The aim of this paper is to generalize results presented
in \cite{LMM:unif} concerning the theory of unification for systems of
equations.

The paper is organized as follows. Section~\ref{sc:basic} contains basic
definitions and notations. First we give a brief overview of syntax and
semantics of first order languages. Then we define positive conjunctive queries
and introduce the important concept of an equivalence relation on queries.
Finally, we introduce finite substitutions and give a short description of some
lattice-theoretic properties of terms.

In Section~\ref{sc:solving} we give an algorithm, called Solved Form Algorithm,
which transforms any query into a solved form. Theorem~\ref{cl:Robinson}
establishes correctness and termination of the algorithm.

In Section~\ref{sc:query} we study lattice-theoretic properties of queries. In
particular, we show that the relationship between two equivalent queries in
solved form is very strong (see Theorem~\ref{cl:query:equiv:order}).

In the last section we present two lattices: the lattice of \EQ{formulas} and
the lattice of finite substitutions. We prove that the both lattices are
isomorphic (see Theorem~\ref{cl:subst:isomorphism}). Finally, we introduce the
notion of application of substitutions to formulas and clarify its relationship
to \EQ{formulas} (see Theorem~\ref{cl:application:correspondence}).

This theory can be regarded as a basis for alternative
presentation of logic programming (see \cite{Komara:logic}).

\section{Notation and definitions}\label{sc:basic}

In this section we recall some basic definitions. We refer to
\cite{Apt:logic,Lloyd:logic,Shoenfield:logic} for a more detailed presentation
of our topics.

\paragraph{Syntax}
The {\em alphabet\/} $L$ for a first order language consists of {\em logical
symbols\/} (a denumerable set of variables, punctuation symbols, connectives
and quantifiers) and two disjoint classes of nonlogical symbols: (i) a set
\FUNC{L} of function symbols (including constants) and (ii) a set
\PRED{L} of predicate symbols. Throughout this paper we assume that the set of
function symbols contains at least one constant. Moreover we always suppose
that the equality symbol $=$ and propositional constants \TRUE\ and \FALSE\ are
contained in all alphabets we shall use. We shall write
\( e_1 \equiv e_2 \)
to denote the syntactical identity of two strings $e_1$ and $e_2$ of symbols.
We denote by \VAR\ the set of all variables.

We use $u$, $v$, $x$, $y$ and $z$, as syntactical variables which vary through
variables; $f$, $g$ and $h$ as syntactical variables which vary through
function symbols; $p$ and $q$ as syntactical variables which vary through
predicate symbols excluding $=$; and $a$, $b$, $c$ and $d$ as syntactical
variables which vary through constants.

The {\em first order language} consists of two classes of strings of symbols
over a given alphabet $L$: (i) a set of terms, denoted \TERM{L}, and (ii) a set
of all well-formed formulas. We use $r$, $s$ and $t$, as syntactical variables
which vary through terms; $F$, $G$ and $H$ as syntactical variables which vary
through formulas.  An {\em equation\/} is a formula of the form $s=t$; and an
{\em atom\/} is a formula of the form
\( p(s_1, \ldots, s_n) \).
We use $A$, $B$ and $C$, as syntactical variables which vary through atoms. A
formula is called {\em positive\/} if it is constructed from propositional
constants \TRUE\ and \FALSE, and from equations and atoms using the conjunction
$\wedge$, the disjunction $\vee$ and quantifiers. By an {\em equational\/}
formula we mean any formula having no occurrence of atoms.

Consider a term $s$. Then \VARS{s} denotes the set of variables appearing in
$s$. If \VARS{s} is empty, then the term $s$ is called {\em ground\/}.
Similarly, \VARS{F} denotes the set of free variables of a formula $F$. $F$ is
said to be {\em closed\/} if
\( \VARS{F} = \emptyset \).
Let
\( x_1, \ldots, x_n \)
be all distinct variables occurring freely in  a formula $F$ in this order. We
write
\( (\forall) F \)
or
\( (\exists) F \)
for
\( (\forall x_1) \ldots (\forall x_n) F \)
or
\( (\exists x_1) \ldots (\exists x_n) F \),
respectively. We call
\( (\forall) F \)
or
\( (\exists) F \)
the {\em universal closure\/} or the {\em existential closure\/} of $F$,
respectively.

In order to avoid the awkward expression {\em the first order language over an
alphabet \LA{}\/} we will simple say {\em the first order language \LA{}\/} (or
{\em the language \LA{}\/} for short).

\paragraph{}
In Section~\ref{subs:subst} we introduce the notion of application of finite
substitutions to formulas. The definition is based on a weaker form of this
application. We follow here \cite{Shoenfield:logic}:
\begin{enumerate}
	\item [(a)] 
			\( t_{x_1, \ldots, x_n}[s_1, \ldots, s_n] \) 
			denotes a term obtained from $t$ by simultaneously replacing of each
			occurrence
			\( x_1, \ldots, s_n \) 
			in $t$ by 
			\( s_1, \ldots, s_n \),
			respectively.
	\item [(b)]
			\( F_{x_1, \ldots, x_n}[s_1, \ldots, s_n] \) 
			denotes a formula obtained from $F$ by simultaneously replacing
			of each free occurrence 
			\( x_1, \ldots, x_n \) 
			in $F$ by 
			\( s_1, \ldots, s_n \),
			respectively.
\end{enumerate}
Whenever
\( t_{x_1, \ldots, x_n}[s_1, \ldots, s_n] \) 
or
\( F_{x_1, \ldots, x_n}[s_1, \ldots, s_n] \) 
appears,
\( x_1, \dots , x_n\) 
are restricted to represent distinct variables. Moreover, in (b) we always
suppose that each term $s_i$ is {\em substitutible\/} for $x_i$ in $F$
i.e. for each variable $y$ occurring in $s_i$, no part of $F$ of the form
\( (\exists y) G \) (or \( (\forall y) G \) ) 
contains an occurrence of $x_i$ which is free in $F$. We shall omit the
subscripts
\( x_1, \ldots, x_n \)
when they occur freely in $F$ in this order and
\( \VARS{F} = \{ x_1, \ldots, x_n \} \).

We say that $F'$ is a {\em variant\/} of $F$, if $F'$ can be obtained
from $F$ by a sequence of replacement of the following type: replace a
part
\( (\exists x) G \) 
or 
\( (\forall x) G \)
by 
\( (\exists y) G_x[y] \) 
or by
\( (\forall x) G_x[y] \),
respectively, where $y$ is a variable not free in $G$.

\paragraph{}
By {\em free equality axioms\/} for a language \LA{} (see
\cite{Apt:logic,Lloyd:logic}), we mean the theory \FEA{L} consisting of the
following formulas:
\begin{enumerate}
	\item[(a)]
			\( f(x_1, \ldots, x_n) = f(y_1, \ldots, y_n) \leftrightarrow 
			x_1 = y_1 \wedge \ldots \wedge x_n = y_n \)
			for each {\em n\/}-ary function symbol $f$,
	\item[(b)]
			\( f(x_1, \ldots, x_n) = g(y_1, \ldots, y_m) \leftrightarrow \FALSE \)
			for each {\em n\/}-ary function symbol $f$ and {\em m\/}-ary function
			symbol $g$ such that 
			\( f \not \equiv g \),
	\item[(c)]
			\( x = t \leftrightarrow \FALSE \)
			for each variable $x$ and term $t$ such that 
			\( x \not \equiv t \) 
			and $x$ occurs in $t$.
\end{enumerate}
Since we identify constants with {\em 0\/}-ary function symbols, then (b)
includes
\( a \not \equiv b \) 
for pairs of distinct constants as a special case.

\paragraph{Semantics}
A {\em pre-interpretation\/} $J$ for a language \LA{}\ consists of (i) a
non-empty universe \UU{L}{J}, called a {\em domain\/} of $J$, and (ii) a fixed
interpretation of all function symbols. The equality $=$ is interpreted as the
identity on \UU{L}{J}.  An {\em interpretation\/} $I$ for \LA{}\ is based on
$J$ (or just {\em J-interpretation} for short) if it is obtained from $J$ by
selecting some interpretation of predicate symbols.

Consider a pre-interpretation $J$. A variable assignment
\( h : \VAR \rightarrow \UU{L}{J} \) 
will be called a {\em valuation\/} over $J$ (or a {\em J-valuation\/} for
short).  The set of all {\em J\/}-valuations is denoted by \VV{L}{J}. Obviously
each {\em J\/}-valuation $h$ has a unique homomorphic extension $h'$ from
\TERM{} into \UU{L}{J}. We shall write $h(s)$, where $s$ is a term, instead of
$h'(s)$. We call $h(s)$ a {\em J-instance\/} of $s$. Consider an atom 
\( A \equiv p(s_1, \ldots, s_n) \)
and a {\em J\/}-valuation $h$. The generalized atom 
\( h(A) \equiv p(h(s_1), \ldots, h(s_n)) \).
will be called a {\em J-instance\/} of $A$. The set of all {\em J\/}-instances
of atoms, called {\em J-base\/}, is denoted by \BB{L}{J}. We shall identify
interpretations based on $J$ with subsets of \BB{L}{J}. We shall use the
notation \CCA{h}{x}{d}, where $x$ is a variable and
\( d \in \UU{L}{J} \),
to denote a valuation defined as follows:
\[
	\CCB{h}{x}{d}(y) =
		\left\{
			\begin{array}{ll}
				d    & \mbox{if \( y \equiv x \)} \\
				h(y) & \mbox{otherwise}
			\end{array}
		\right.
\]

\begin{example}\label{ex:herbrand}
	The {\em Herbrand\/} pre-interpretation \HA\ for $L$ is defined as follows:
	\begin{enumerate}
		\item[(a)]
				Its domain is the set \UU{L}{\HB} of all ground terms of $L$;
				called the {\em Herbrand universe\/}.
		\item[(b)]
				Each constant in $L$ is assigned to itself.
		\item[(c)]
				If $f$ is an {\em n\/}-ary function symbol in $L$ then it is
				assigned to the mapping from $(U_L^{\HB})^n$ to \UU{L}{\HB} defined
				by assigning the ground term
				\( f(s_1, \ldots, s_n) \)
				to the sequence
				\( s_1, \ldots, s_n \)
				of ground terms.
	\end{enumerate}
	By a {\em Herbrand\/} interpretation for $L$ we mean any interpretation
	based on \HA. As remarked above we shall identify Herbrand interpretations
	with subsets of the set \BB{L}{\HB} called the {\em Herbrand base\/}.
\end{example}

Given a formula $F$ we define its {\em truth\/} in a {\em J\/}-valuation $h$
and an interpretation $I$ based on $J$, written as \TRUTH{I}{h}{F}, in the
obvious way. In particular, \TRUTH{I}{h}{s = t} iff
\( h(s) = h(t) \).
We call
\( h \in \VV{L}{J} \)
a {\em solution\/} of $F$ in $I$, if \TRUTH{I}{h}{F}. The set of solutions of
$F$ in $I$ is denoted by \ANS{F}{I}{J}. So:
\[	\ANS{F}{I}{J} = \{ h \in \VV{L}{J} \mid \TRUTH{I}{h}{F} \} \]
We say that a formula $F$ is {\em true\/} in $I$, written as \TRUTH{I}{}{F},
when for all valuations
\( h \in \VV{L}{J} \), 
\TRUTH{I}{h}{F}. We say that a formula is {\em valid\/} when it is true in any
interpretation. 

Consider a theory $T$ over \LA{}. An interpretation $I$ is a {\em model\/} for
$T$ if \TRUTH{I}{}{F} for any $F$ from $T$. A formula $F$ (over \LA{}) is  a
{\em logical consequence\/} of $T$ if any model $I$ for $T$ is a model for $F$,
as well. We write \TRUTH{T}{}{F} in \LA{} (or \TRUTH{T}{}{F} for short).

Consider an equational formula $F$. Note that the semantics of $F$ depends only
on an interpretation of function symbols. Consequently, given a
pre-interpretation $J$, the truth of $F$ in a {\em J\/}-valuation $h$ is
well-defined. We shall write \TRUTH{J}{h}{F}. We call
\( h \in \VV{L}{J} \)
a solution of $F$ in $J$ (or {\em J-solution\/} of $F$ for short), if
\TRUTH{J}{h}{F}. The set of {\em J\/}-solutions of $F$ is denoted by
\SOLN{J}{F}. So:
\[ \SOLN{J}{F} = \{ h \in \VV{L}{J} \mid \TRUTH{J}{h}{F} \} \]
$F$ is said to be {\em true\/} in $J$, written as \TRUTH{J}{}{F}, when for all
valuations
\( h \in \VV{L}{J} \), 
is \TRUTH{J}{h}{F}. We say that $J$ is a model for \FEA{L}, when any axiom of
\FEA{L} is true in $J$.

The following theorems will be used in the sequel (see
\cite{Shoenfield:logic}).

\begin{theorem}[Variant Theorem]\label{cl:variant}
	If $F'$ is a variant of $F$, then 
	\( F' \leftrightarrow F \)
	is valid.
\end{theorem}

\begin{theorem}[Theorem on Constants]\label{cl:constants}
	Let $T$ be a theory and $F$ a formula. If 
	\( a_1, \dots, a_n \) 
	are distinct constants not occurring in $T$ and $F$, then 
	\[ \TRUTH{T}{}{F}\ {\rm iff}\ \TRUTH{T}{}{F[a_1, \ldots, a_n]} \]
\end{theorem}

\begin{theorem}\label{cl:basic}
	Consider a theory $T$ and formulas $F$ and $G$. Then:
	\begin{enumerate}
		\item \( F \rightarrow (\exists x) F \)
				is valid.
		\item If
				\TRUTH{T}{}{F \rightarrow G},
				then
				\TRUTH{T}{}{(\exists x) F \rightarrow G}
				provided $x$ is not free in $G$.
		\item If
				\TRUTH{T}{}{F \rightarrow G},
				then
				\TRUTH{T}{}{(\exists x) F \rightarrow (\exists x) G}
	\end{enumerate}
\end{theorem}

\paragraph{Queries}

A {\em conjunctive query\/} is a formula constructed from propositional
constants, equations and atoms using the conjunction $\wedge$, the negation
$\neg$ and the existential quantifier $\exists$. In this paper we deal only
with conjunctive queries which are positive formulas. From now by a query we
mean a positive conjunctive query. We denote by \CARD{Q} the number of atoms
occurring in $Q$. If
\( \CARD{Q} = 0 \),
then $Q$ is called an {\em empty\/} query.

A query $Q$ is said to be in {\em solved form\/} if 
\begin{itemize}
	\item $Q$ is a propositional constant \TRUE\ or \FALSE\ or
	\item $Q$ is of the form
			\[ (\exists z_1) \ldots (\exists z_k) (x_1 = s_1 \wedge \ldots \wedge 
				x_n = s_n \wedge A_1 \wedge \ldots \wedge A_m), \]
			where (i) $x_i$'s and $z_j$'s are distinct variables, (ii) $x_i$'s
			occur nor in the right hand side of any equation nor in any atom,
			(iii) each $z_j$ has at least one occurrence in the conjunction and 
			(iv) 
			\( z_j \not \equiv s_i \) 
			for any $i$ and $j$.
\end{itemize}
The variables 
\( x_1, \ldots, x_n \) 
are said to be {\em eliminable} and the set 
\( \{ x_1, \ldots, x_n \} \) 
is denoted by \ELIM{Q}. The remaining free variables in $Q$ are called {\em
parameters} and the set of parameters is denoted by \PARAM{Q}. So 
\( \VARS{Q} = \ELIM{Q} \cup \PARAM{Q} \). 
The set 
\( \{ z_1, \ldots, z_k \} \) 
of (existentially) bound variables in $Q$ is denoted by \BOUND{Q}. We put 
\( \ELIM{\TRUE} = \PARAM{\TRUE} = \BOUND{\TRUE} = \emptyset \).

In Section~\ref{sc:solving} we shall describe an algorithm, called Solved Form
Algorithm, which transforms every query into a solved form. The algorithm is
sound in the following sense: if $Q'$ is a computed solved form of a query $Q$,
then $Q'$ is semantically equivalent to $Q$ w.r.t. \FEA{\LB{}} i.e.
\begin{equation}\label{eq:qrs:1}
	\TRUTH{\FEA{\LB{}}}{}{Q' \leftrightarrow Q}
\end{equation}
Actually, we prove that Solved Form Algorithm has a stronger property than 
(\ref{eq:qrs:1}). In particular, as we see later, the algorithm preserves the
number of atoms. Consequently,
\( \CARD{Q'} = \CARD{Q} \).
For this purpose we introduce a modified concept of solution sets for queries.
Instead of the mapping
\( \ANS{Q}{}{J} : 2^{B_L^J} \rightarrow \VV{L}{J} \),
which has a single interpretation as an argument, we shall consider a mapping 
\( \MANS{Q}{}{J} : 2^{B_L^J} \times \ldots \times 2^{B_L^J}
\rightarrow \VV{L}{J} \)
defined on sequences \OLA{I}{} of interpretations. Each component in \OLA{I}{}
will serve as an ``input'' for the corresponding atom in $Q$. We follow here
ideas developed in \cite{SK:lop}.

By a {\em multiinterpretation\/} based on a pre-interpretation $J$ we mean any
finite sequences of interpretations based on $J$. The empty sequence is denoted
by \EMPTYSEQA. We shall use overlined letters to denote multiinterpretations.
We write
\( \OLB{I}{_1} \OLB{I}{_2} \)
for the concatenation of multiinterpretations \OLA{I}{_1} and \OLA{I}{_2}. The
length of \OLA{I}{} is denoted by \CARD{\OLB{I}{}}. We say that
\OLA{I}{} {\em is for\/} a query $Q$ if
\( \CARD{\OLB{I}{}} = \CARD{Q} \).

Consider a pre-interpretation $J$, a query $Q$ and a multiinterpretation
\OLA{I}{} for $Q$ based on $J$. The set of solutions of the query $Q$ in
\OLA{I}{} is a set of $J$-valuations, denoted by \MANS{Q}{\OLB{I}{}}{J}, and it
is defined inductively as follows:
\begin{enumerate}
	\item \( \MANS{\TRUE}{\EMPTYSEQB}{J} = \VV{L}{J} \)
			and
			\( \MANS{\FALSE}{\EMPTYSEQB}{J} = \emptyset \)
	\item \( \MANS{s = t}{\EMPTYSEQB}{J} = \SOLN{J}{s = t} \)
	\item \( \MANS{A}{I}{J} = \ANS{A}{I}{J} \),
			if $A$ is an atom.
	\item \( \MANS{Q_1 \wedge Q_2}{\OLB{I}{_1}\OLB{I}{_2}}{J}  = 
				\MANS{Q_1}{\OLB{I}{_1}}{J} \cap \MANS{Q_2}{\OLB{I}{_2}}{J} \),
			where 
			\( \CARD{\OLB{I}{_1}} = \CARD{Q_1} \)
			and 
			\( \CARD{\OLB{I}{_2}} = \CARD{Q_2} \)
	\item \( \MANS{(\exists x)Q}{\OLB{I}{}}{J} = 
				\{ h \in \VV{L}{J} \mid 
					\CCB{h}{x}{d} \in \MANS{Q}{\OLB{I}{}}{J}\ \ 
					{\rm for\ some}\ d \in \UU{L}{J} \} \)
\end{enumerate}
We call a $J$-valuation $h$ from \MANS{Q}{\OLB{I}{}}{J} a {\em solution\/} of
$Q$ in \OLA{I}{}. Note that
\( \ANS{Q}{I}{J} = \MANS{Q}{\OLB{I}{}}{J} \), 
where
\( \OLB{I}{} = I, \ldots, I \)
and
\( \CARD{\OLB{I}{}} = \CARD{Q} \).

Consider a pre-interpretation $J$. Then a preorder relation $\preceq_J$ and an
equivalence relation $\approx_J$ on queries is defined as follows:
\begin{enumerate}
	\item[(a)]
			\( Q \preceq_J Q' \)
			iff
			\( \CARD{Q} = \CARD{Q'} \)
			and
			\( \MANS{Q}{\OLB{I}{}}{J} \subseteq \MANS{Q'}{\OLB{I}{}}{J} \)
			for any multiinterpretations \OLA{I}{} based on $J$
	\item[(b)]
			\( Q \approx_J Q' \)
			iff
			\( \CARD{Q} = \CARD{Q'} \)
			and
			\( \MANS{Q}{\OLB{I}{}}{J} = \MANS{Q'}{\OLB{I}{}}{J} \)
			for any multiinterpretations \OLA{I}{} based on $J$
\end{enumerate}
Finally, we say 
\begin{enumerate}
	\item[(a)]
			$Q'$ is {\em more general\/} than $Q$, written as 
			\( Q \preceq Q' \),
			if for any pre-interpretation $J$, which is a model for \FEA{\LB{}}, 
			we have
			\( Q \preceq_J Q' \)
	\item[(b)]
			$Q'$ is {\em equivalent\/} to $Q$, written as 
			\( Q \preceq Q' \),
			if 
			\( Q \preceq Q' \)
			and
			\( Q' \preceq Q \)
\end{enumerate}
Then the precise form of soundness of Solved Form Algorithm is following (see
Theorem~\ref{cl:Robinson}): {\em the solved form algorithm applied to a query
$Q$ will return an equivalent one in solved form.}

We say that a query $Q$ is {\em consistent\/} if there is an interpretation $I$
(based on a pre-interpretation $J$), which is a model for \FEA{\LB{}}, such
that \TRUTH{I}{h}{Q} for some
\( h \in \VV{\LB{}}{J} \),
or equivalently if \NOTTRUTH{\FEA{\LB{}}}{}{\neg Q}. Note that the consistency
of queries in solved form can be checked directly: if $Q$ is a query in solved
form, then $Q$ is consistent iff
\( Q \not \equiv \FALSE \).
As we marked above Solved Form Algorithm preserves equivalence and
consequently it preserves consistency, as well. Hence the consistency of a
query $Q$ can be reached directly by the the form of its (computed) solved
form.

\paragraph{Finite Substitutions}
By a {\em finite substitution\/} we mean any mapping of terms to variables from
a finite set $X$ of variables. Let
\( X = \{ x_1, \ldots, x_n \} \).
We shall use the standard set-theoretic notation 
\( \sigma = \{ \BINDA{x_1}{s_1}, \ldots, \BINDA{x_n}{s_n} \} \) 
to denote $\sigma$, where
\( s_i \equiv \sigma(x_i) \).
We say that $\sigma$ is {\em over\/} $X$. We denote by \DOM{\sigma} or
\RANGE{\sigma} the set $X$ or the set of variables occurring in terms
\( s_1, \ldots, s_n \),
respectively. The pair \BINDA{x}{s} is called a {\em binding}. In the sequel by
a substitution we always mean a finite substitution.

Let $\sigma$ and $\theta$ are substitutions with disjoint domains. By
\( \sigma \cup \theta \)
we mean a substitution over
\( \DOM{\sigma} \cup \DOM{\theta} \)
assigning
\( \sigma(x) \)
or
\( \theta(x) \)
to
\( x \in \DOM{\sigma} \)
or
\( x \in \DOM{\theta} \),
respectively. We call $\sigma$ a {\em permutation\/} if it is one-to-one
mapping from \DOM{\sigma} onto \RANGE{\sigma}. There is unique substitutions
\SUBEMPTYA\ with the empty domains; it will be called the {\em empty\/}
substitution.

A substitution
\( \sigma = \{ \BINDA{x_1}{s_1}, \ldots, \BINDA{x_n}{s_n} \} \)
is {\em applicable\/} to a term $t$ if \DOM{\sigma}
contains all variables occurring in $t$. Then an {\em application\/} of
$\sigma$ to $t$, denoted $t \sigma$,
is defined as the term
\( t_{x_1, \ldots, x_n}[s_1, \ldots, s_n] \).
The term $t \sigma$ is called an {\em instance of\/} $t$. If $\sigma$ is a
permutation, then $t \sigma$ and $t$ are said to be {\em variants\/}.

A substitution
\( \sigma = \{ \BINDA{x_1}{s_n}, \ldots, \BINDA{x_n}{s_n} \} \)
is {\em applicable\/} to a formula $F$ if each $s_i$ is substitutible for $x_i$
in $F$. Then an {\em application\/} of $\sigma$ to $F$, denoted
\( F \sigma \),
is defined as a formula
\( F_{x_1, \ldots, x_n}[s_1, \ldots, s_n] \).
In Section~\ref{subs:subst} we generalize this operation for arbitrary
substitutions.

A substitution $\theta$ is {\em applicable\/} to $\sigma$ if 
\( \RANGE{\sigma} \subseteq \DOM{\theta} \).
Then a {\em composition\/} of $\sigma$ and $\theta$ is a substitution over
\DOM{\sigma}, denoted
\( \sigma \theta \),
and it is defined in the obvious way. If $\theta$ is a permutation, then we say
that $\sigma$ and $\sigma \theta$ are {\em variants\/}.

We denote by \RESTR{\sigma}{X} a {\em restriction\/} of $\sigma$ onto $X$. So
\RESTR{\sigma}{X} is a substitution over
\( \DOM{\sigma} \cap X \).
We call $\sigma$ an {\em extension\/} of \RESTR{\sigma}{X} onto \DOM{\sigma}.

\paragraph{Terms}
In Section~\ref{sc:query} we shall study lattice-theoretic properties of
queries. To prove the claims of that section we need the following simple
results concerning terms.

Fist we introduce the preorder relation $\preceq$ and the equivalence relation
$\approx$ of terms:

\begin{itemize}
	\item A term $t$ is {\em more general\/} than a term $s$, 
			written as
			\( s \preceq t \), 
			if there is a substitution $\theta$ such that 
			\( s \equiv t \theta \).
	\item A term $t$ is {\em equivalent\/} to a term $s$, written as
			\( s \approx t \), 
			if 
			\( s \preceq t \) 
			and 
			\( t \preceq s \).
\end{itemize}

On the other hand each pre-interpretation $J$ defines a preorder $\preceq_J$
and an equivalence $\approx_J$ as follows. Let \INST{J}{s} denote the set of
all $J$-instances of the term $s$. Then:
\begin{itemize}
	\item A term $t$ is {\em more general\/} w.r.t. $J$ than a term $s$, 
			written as 
			\( s \preceq_J t \), 
			if 
			\( \INST{J}{s} \subseteq \INST{J}{t} \).
	\item A term $t$ is {\em equivalent\/} w.r.t. $J$ (or {\em $J$-equivalent\/} 
			for short) to a term $s$, written as 
			\( s \approx_J t \), 
			if 
			\( \INST{J}{s} = \INST{J}{t} \).
\end{itemize}
Clearly if
\( \sigma \preceq \theta \)
or
\( \sigma \approx \theta \)
then
\( \sigma \preceq_J \theta \)
or
\( \sigma \approx_J \theta \),
respectively. The following example shows that $\preceq_J$ does not coincide
with $\preceq$ in general.

\begin{example}\label{ex:terms:trivial}
	Suppose that the Herbrand universe for $L$ consists of the constant $a$
	only. Let $x$ be a variable. Then 
	\( x \preceq_{\HB} a \) 
	and 
	\( x \not \preceq a \).
\end{example}

Proposition~\ref{cl:terms:preorders} states that both preorders $\preceq_J$ and
$\preceq$ are identical (and hence 
\( \approx_J\ \equiv\ \approx \)
as well), provided that $J$ satisfies the following two conditions: (i) $J$ is
a model for \FEA{L} and (ii) its domain has at least two elements (such
pre-interpretations will be called {\em non-trivial}). The next simple lemma
will be used in the sequel.

\begin{lemma}\label{cl:terms:nontrivial}
	Suppose that $J$ is a non-trivial model of \FEA{L}. Let $s$ and $t$ be both
	non-variable terms.
	\begin{enumerate}
		\item \( \INST{J}{s} \not = \VV{L}{J} \).
		\item If $s$ and $t$ have different principal functors, then 
				\( \INST{J}{s} \cap \INST{J}{t} = \emptyset \).
	\end{enumerate}
\end{lemma}

{\bf Proof:}
Straightforward.
\END

\begin{proposition}\label{cl:terms:preorders}
	Let $s$ and $t$ be terms. If $J$ is a non-trivial model of \FEA{L}, then:
	\begin{enumerate}
		\item \( s \preceq_J t\ \ {\rm iff}\ \ s \preceq t \)
		\item \( s \approx_J t\ \ {\rm iff}\ \ s \approx t \) 
				iff $s$ and $t$ are variants
	\end{enumerate}
\end{proposition}

To prove the claim we first introduce the notion of difference sets for terms.
By a {\em difference set\/} of terms $s$ and $t$ we mean a set of pairs of
terms, denoted
\DIFF{s}{t}, which is defined inductively as follows:

\begin{itemize}
	\item \( \DIFF{x}{x} = \{ (x,x) \} \)
	\item \( \DIFF{a}{a} = \emptyset \)
	\item \( \DIFF{f(s_1, \ldots, s_n)}{f(t_1, \ldots, t_n)} = 
			\DIFF{s_1}{t_1} \cup \ldots \cup \DIFF{s_n}{t_n} \) 
	\item \( \DIFF{s}{t} = \{ (s, t) \} \) otherwise
\end{itemize}

\noindent
For example 
\[ \DIFF{f(x,g(z),z)}{f(y,a,z)} = \{ (x,y), (g(z),a), (z,z) \} \]
Difference sets have the following property: 

\begin{lemma}\label{cl:terms:diff}
	Let $s$ and $t$ be terms. Suppose that $J$ is a model of \FEA{L}. If $h$ 
	and $g$ are $J$-valuations such that 
	\( h(s) = g(t) \), 
	then 
	\( h(s') = g(t') \) 
	for any pair 
	\( (s',t') \in \DIFF{s}{t} \). 
\end{lemma}

{\bf Proof:}
Straightforward.
\END
\medskip

\noindent
Moreover notice that 

\begin{equation}\label{eq:terms:diff}
	\DIFF{s}{t \theta} = \bigcup \{ \DIFF{s'}{t' \theta} \mid (s',t') \in 
	\DIFF{s}{t} \}, 
\end{equation}

\noindent
for any substitution $\theta$ applicable to $t$. Now we are ready to prove the
desired proposition.

\medskip

{\bf Proof of Proposition~\ref{cl:terms:preorders}:}
First we prove that 
\( s \preceq_J t \), 
implies 
\( s \preceq t \). 

Let 
\( (s',t') \in \DIFF{s}{t} \). 
Then $t'$ must be a variable, since otherwise by
Lemma~\ref{cl:terms:nontrivial} there is
\( h \in \VV{L}{J} \) 
such that 
\( h(s') \not \in \INST{J}{t'} \). 
Then Lemma~\ref{cl:terms:diff} implies that 
\( h(s) \not \in \INST{J}{t} \). 
Contradiction.

Assume that 
\( (r_1,x) \in \DIFF{s}{t} \) 
and 
\( (r_2,x) \in \DIFF{s}{t} \). 
Then terms $r_1$ and $r_2$ must be the same, since otherwise there is 
\( h \in \VV{L}{J} \) 
such that 
\( h(r_1) \not = h(r_2) \). 
Because 
\( \INST{J}{s} \subseteq \INST{J}{t} \), 
there is 
\( g \in \VV{L}{J} \) 
such that 
\( h(s) = g(t) \). 
Lemma~\ref{cl:terms:diff} implies that 
\( h(r_1) = g(x) \) 
and 
\( h(r_2) = g(x) \) 
Contradiction.

Now let $\theta$ be a substitution over \VARS{t} such that for any 
\( x \in \VARS{t} \) 
it holds 
\( \theta(x) \equiv r \), 
where 
\( (r,x) \in \DIFF{s}{t} \). 
$\theta$ is well defined since of the form \DIFF{s}{t}. From
(\ref{eq:terms:diff}) 
we have 
\( \DIFF{s}{t \theta} = \bigcup \{ \DIFF{x}{x} \mid x \in \VARS{s} \} \). 
Hence 
\( s \equiv t \theta \) i.e.
\( s \preceq t \). 

If 
\( s \approx_J t \) 
or 
\( s \approx t \), 
then the substitution $\theta$ is a permutation and hence $s$ and $t$ are
variants. 
\END

\paragraph{Canonical language}
In Section~\ref{sc:eqnsubst} we shall prove the so-called Compactness Theorem.
It relies on Proposition~\ref{cl:canonical}.

We say that the first order language \LA{2} is an {\em extension\/} of a
first order language \LA{1} if every nonlogical symbol of \LA{1} is a
nonlogical symbol of \LA{2}. The following claim will be used in the proof of 
Proposition~\ref{cl:canonical}.

\begin{lemma}[\cite{Shoenfield:logic}]\label{cl:extension}
	Consider a theory $T$ over \LA{1}. If \LA{2} is obtained from \LA{1} by
	adding some constants, then
	\[ \TRUTH{T}{}{F}\ {\rm in}\ \LB{1}\ \ {\rm iff}\ \ 
		\TRUTH{T}{}{F}\ {\rm in}\ \LB{2} \]
	for every formula $F$ over \LA{1}.
\end{lemma}

\noindent
We restrict our attention only to extensions which are obtained by adding
denumerable number of new constants. We say that a first order language
\LA{c} is a {\em canonical language\/} for a language \LA{} if \LA{c} is
obtained by adding denumerable number of new constants to \LA{}.

\begin{proposition}\label{cl:canonical}
	Let \LA{c} be a canonical language for \LA{} and \HA\ a Herbrand
	pre-in\-ter\-pre\-ta\-tion for \LA{c}. Then 
	\[ \TRUTH{\HB}{}{F}\ \ {\rm iff}\ \ \TRUTH{\FEA{\LB{}}}{}{F}\ {\rm in}\ 
	\LB{} \]
	for every equational formula $F$ over \LA{}.
\end{proposition}

{\bf Proof:}
Lemma~\ref{cl:extension} implies that it is sufficient to show that
\begin{equation}\label{eq:canonical:1}
	\TRUTH{\HB}{}{F}\ \ {\rm iff}\ \ \TRUTH{\FEA{\LB{}}}{}{F}\ {\rm in}\ \LB{c}
\end{equation}
holds for every equational formula $F$ over \LA{}. The proof of IF part in
(\ref{eq:canonical:1}) is based on the following claim:
\begin{equation}\label{eq:canonical:2}
	{\rm if}\ \TRUTH{\HB}{}{F},\ {\rm then}\ \ \TRUTH{\FEA{\LB{}}}{}{F}\ 
	{\rm in}\ \LB{c}
\end{equation}
for every closed equational formula $F$ over \LA{c}. We prove
(\ref{eq:canonical:2}) by induction on closed formulas.

If 
\( \TRUTH{\HB}{}{s = t} \), 
where $s$ and $t$ are ground terms, then 
\( s \equiv t \). 
Thus 
\( \TRUTH{\FEA{\LB{}}}{}{s = t} \) 
in \LA{c}. By a straightforward application of the induction hypothesis we
prove (\ref{eq:canonical:2}) for a formula $F$ of the form
\( F \equiv \neg G \),
\( F \equiv G \wedge H \),
\( F \equiv G \vee H \),
\( F \equiv G \rightarrow H \) 
or 
\( F \equiv G \leftrightarrow H \),
respectively.

Suppose 
\( F \equiv (\exists x) G \). 
If 
\( \TRUTH{\HB}{}{F} \), 
then 
\( \TRUTH{\HB}{}{G_x[s]} \) 
for some ground term $s$. By induction hypothesis we have 
\( \TRUTH{\FEA{\LB{}}}{}{G_x[s]} \) 
in \LA{c}. Since 
\( G_x[s] \rightarrow (\exists x) G \) 
is valid, we have 
\( \TRUTH{\FEA{\LB{}}}{}{F} \) 
in \LA{c}.

Assume finally that 
\( F \equiv (\forall x) G \). 
If 
\( \TRUTH{\HB}{}{F} \), 
then 
\( \TRUTH{\HB}{}{G_x[s]} \)  
for any ground term $s$. By induction hypothesis we have for any 
\( s \in \UU{\LB{}}{\HB} \) 
that 
\( \TRUTH{\FEA{\LB{}}}{}{G_x[s]} \) 
in \LA{c}. If $s$ is a constant $a$ occurring nor in \FEA{\LB{}} and nor in
$G$, then by Theorem~\ref{cl:constants} we have
\( \TRUTH{\FEA{\LB{}}}{}{G} \) 
in \LA{c} and therefore
\( \TRUTH{\FEA{\LB{}}}{}{F} \) 
in \LA{c}.

Now let $F$ be arbitrary (not necessary closed) equational formula over \LA{}.
Consider distinct
\( a_1, \ldots, a_n \) 
constants not occurring in \FEA{\LB{}} and $F$. If \TRUTH{\HB}{}{F}, then 
\TRUTH{\HB}{}{F[a_1, \ldots, a_n]} and hence by (\ref{eq:canonical:2}) we have 
\TRUTH{\FEA{\LB{}}}{}{F[a_1, \ldots, a_n]} in \LA{c}.
Theorem~\ref{cl:constants} implies that
\TRUTH{\FEA{\LB{}}}{}{F} in \LA{c}. This concludes the proof of IF part in
(\ref{eq:canonical:1}).

The proof of ONLY-IF part in (\ref{eq:canonical:1}) is straightforward.
\END

\section{Solving queries}\label{sc:solving}

In this section we present an algorithm, called Solved Form Algorithm, which
transforms any query into an equivalent one in solved form. We follow here the
presentation of \cite{LMM:unif}.

\paragraph{Solved Form Algorithm}
For a given query $Q$ non-deterministically apply the following elementary
steps (1) - (12). We write
\( (\exists \OLB{y}{}) \)
instead of
\( (\exists y_1) \ldots (\exists y_k) \),
if \OLA{y}{} is the sequence
\( y_1, \ldots, y_k \).

The first group of elementary actions is determined by the form of a selected
equation in $Q$.
\begin{enumerate}
	\item[(1)]
			\( f(s_1, \ldots, s_n) = f(t_1, \ldots, t_n) \)
			replace by
			\( s_1 = t_1 \wedge \ldots \wedge s_n = t_n \)
	\item[(2)]
			\( f(s_1, \ldots, s_n) = g(t_1, \ldots, t_m) \)
			replace by \FALSE, if $f$ and $g$ are distinct symbols
	\item[(3)]
			\( x = t \)
			replace by \FALSE, if $x$ and $t$ are distinct terms such that $x$ 
			occurs in $t$
	\item[(4)]
			\( x = x \)
			replace by \TRUE
\end{enumerate}
The following two actions eliminate a variable $x$ if the selected equation is
of the form $x=t$, where
\( x \not \equiv t \)
and $x$ does not occur in $t$. We suppose that $x=t$ is ''surrounded'' only by
atoms and equations i.e. there is a subquery
\( (\exists \OLB{y}{}) (Q' \wedge x = t \wedge Q'') \)
of the query $Q$, where $Q'$ and $Q''$ are conjunctions only of equations and
atoms. The third action redirects $t=x$ according the form of $t$.
\begin{enumerate}
	\item[(5)]
			\( (\exists \overline{y}) (Q' \wedge x = t \wedge Q'') \)
			replace by
			\( (\exists \overline{y}) (Q'_x[t] \wedge x = t \wedge Q''_x[t]) \),
			if $x$ is not in \OLA{y}{} and it has another (free) occurrence in 
			$Q'$ or in $Q''$
	\item[(6)]
			\( (\exists \overline{y}) (Q' \wedge x = t \wedge Q'') \)
			replace by
			\( (\exists \overline{y}) (Q'_x[t] \wedge \TRUE \wedge Q''_x[t]) \),
			if $x$ is in \OLA{y}{}
	\item[(7)]
			\( (\exists \overline{y}) (Q' \wedge t = x \wedge Q'') \)
			replace by
			\( (\exists \overline{y}) (Q' \wedge x = t \wedge Q'') \),
			if (i) $t$ is not a variable or (ii) $t$ is a distinct variable from 
			$x$ not occurring in \OLA{y}{} and $x$ is in \OLA{y}{}
\end{enumerate}
The operations for eliminating quantifiers are defined as follows:
\begin{enumerate}
	\item[(8)]
			\( (\exists \OLB{y}{'}) (\exists y) (\exists \OLB{y}{''}) Q \)
			replace by
			\( (\exists \OLB{y}{'}) (\exists \OLB{y}{''}) Q \),
			if $y$ is not free in 
			\( (\exists \OLB{y}{''}) Q \)
	\item[(9)]
			\( (\exists \OLB{y}{_1}) Q_1 \wedge (\exists \OLB{y}{_2}) Q_2 \)
			replace by
			\( (\exists \OLB{z}{_1}) (\exists \OLB{z}{_2}) (R_1 \wedge 
				R_2) \),
			where 
			\( (\exists \OLB{z}{_i}) R_i \)
			is a variant of 
			\( (\exists \OLB{y}{_i}) Q_i \) for $i = 1,2$ such that 
			\( \VARS{R_1} \cap \OLB{z}{_2} = \emptyset \)
			and 
			\( \VARS{R_2} \cap \OLB{z}{_1} = \emptyset \);
			we suppose that 
			\( \OLB{y}{_1} \not = \emptyset \)
			or 
			\( \OLB{y}{_2} \not = \emptyset \) 
\end{enumerate}
Finally, we have:
\begin{enumerate}
	\item[(10)]
			\( A \wedge s = t \)
			replace by
			\( s = t \wedge A \),
			where $A$ is an atom.
	\item[(11)]
			delete any occurrence of the propositional constant \TRUE
	\item[(12)]
			replace $Q$ by \FALSE, if $Q$ obtains at least one occurrence of the 
			propositional constant \FALSE
\end{enumerate}
The algorithm terminates with $Q'$ as the output when no step can be applied to
$Q'$ or when \FALSE\ has been returned. We write \REWRA{Q}{}{Q'} if $Q'$ can be
obtained from $Q$ by one step. By \REWRB{}{} we mean the reflexive and
transitive closure of \REWRA{}{}{}.

\paragraph{}
The following theorem establishes the correctness and the termination of the
solved form algorithm.

\begin{theorem}\label{cl:Robinson}
	The solved form algorithm applied to a query will return an equivalent one
	in solved form after finite number of steps.
\end{theorem}

{\bf Proof:}
We only outlined the proof (see \cite{LMM:unif} for details). {\em
Correctness\/}. Let $Q'$ is obtained from $Q$ using the step (1) -- (5) or (7)
-- (12). Clearly
\( Q' \approx Q \).
The correctness of the step (6) is a trivial consequence of the following
claim: {\em if $x$ does not occur in $s$, then
\( (\exists x) (x = s \wedge F) \leftrightarrow F_x[s] \)
is valid formula\/} (see \cite{Shoenfield:logic}). {\em Termination\/}.
Straightforward by using the same arguments as in \cite{LMM:unif}.
\END

\section{The lattice of queries}\label{sc:query}

In this section we investigate lattice--theoretic properties of queries. In
particular, we will see in Theorem~\ref{cl:query:equiv:order} that the
relationship between two equivalent queries in solved form is very strong. The
theorem states also that
\( \preceq_J\ \equiv\ \preceq \) 
and 
\( \approx_J\ \equiv\ \approx \) 
provided that $J$ is a non-trivial model for \FEA{L}. Finally we show that the
preorder $\preceq$ is well-founded (see Theorem~\ref{cl:query:finite}) i.e.
there is no infinite increasing sequence
\( Q_0 \prec Q_1 \prec Q_2 \prec \ldots \)
of queries. We follow here the presentation of \cite{LMM:unif}.

To this purpose we first show how from a given query in solved form it is
possible to obtain new queries again in solved form. Consider a query
\[ Q \equiv (\exists z_1) \ldots (\exists z_m) (x_1 = s_1  \wedge \ldots x_n
= s_n \wedge A_1 \wedge \ldots A_q) \] 
in solved form. A query $Q'$ is obtained from $Q$ (a) by permutation of
equations and bound variables, (b) by renaming of bound variables or (c) by
redirecting equations if
\begin{enumerate}
	\item[(a)]
			\( Q' \equiv (\exists z_{\rho(1)}) \ldots (\exists z_{\rho(m)}) 
			(x_{\pi(1)} = s_{\pi(1)}  \wedge \ldots \wedge x_{\pi(n)} = 
			s_{\pi(n)} \wedge A_1 \wedge \ldots A_q) \), 
			where 
			$\rho$ or $\pi$ is a permutation of the set 
			\( \{ 1, 2, \ldots, m \} \)
			or 
			\( \{ 1, 2, \ldots, n \} \), 
			respectively.
	\item[(b)]
			$Q'$ is a variant of $Q$.
	\item[(c)]
			\( Q' \equiv Q_{u_1, \ldots, u_k, v_1, \ldots, v_k} [v_1, \ldots, v_k,
			u_1, \ldots, u_k] \),
			where 
			\( u_1 = v_1, \ldots, u_k = v_k \) 
			are equations in $Q$ such that $v_i$'s are distinct variables.
\end{enumerate}
Clearly $Q'$ is also in solved form as well. Using
Theorem~\ref{cl:variant} it is easy to prove that
\( Q \leftrightarrow Q' \) 
is a valid formula. As in \cite{LMM:unif} we define the binary relation ``to be
isomorphic'' on queries in solved form. We say that $Q'$ is {\em isomorphic\/}
to $Q$, where $Q$ is in solved form, if $Q'$ can be obtained from $Q$ by a
step-by-step transformation of the following type: (a) permutation of equations
and bound variables, (b) renaming of bound variables or (c) redirecting
equations. This definition implies the following simple result.

\begin{proposition}\label{cl:query:isomorphism}
	Consider a query $Q$ in solved form. Let $Q'$ be isomorphic to $Q$.
	\begin{enumerate}
		\item $Q'$ is in solved form, as well.
		\item \( Q \leftrightarrow Q' \) is valid.
		\item If $Q$ is consistent, then
				\( \CARD{\ELIM{Q'}} = \CARD{\ELIM{Q}} \),
				\( \CARD{\PARAM{Q'}} = \CARD{\PARAM{Q}} \) 
				and 
				\( \VARS{Q'} = \VARS{Q} \)
	\end{enumerate}
\end{proposition}

Consequently, the relation ``to be isomorphic'' is an equivalence on queries in
solved form. Now we establish preliminary results as in \cite{LMM:unif}. We
assume in the following lemmas that $J$ is a non-trivial model of \FEA{L}.

\begin{lemma}\label{cl:query:elim:1}
	Let $Q$ and $Q'$ be consistent queries in solved form. If 
	\( Q \preceq_J Q' \), 
	then 
	\( \CARD{\ELIM{Q}} \geq \CARD{\ELIM{Q'}} \). 
\end{lemma}

{\bf Proof:} 
To prove the claim we will establish an injective mapping from
\ELIM{Q'} into \ELIM{Q}. 
If $x$ belongs to both sets then $x$ is assigned to itself. Let
\( x \in \ELIM{Q'} \setminus \ELIM{Q} \). 
Then there is an equation 
\( x = t \) 
in $Q'$. As 
\( x \not \in \ELIM{Q} \), 
the term $t$ must be a variable $y$. Really, suppose $t$ is not a variable. Put
\( \OLB{I}{} = \BB{L}{J}, \ldots, \BB{L}{J} \),
where
\( \CARD{\OLB{I}{}} = \CARD{Q} = \CARD{Q'} \).
Let $d$ be an element from \UU{L}{J} distinct from all $J$-instances of the
term $t$. Then there is a solution $h$ of $Q$ in \OLA{I}{} such that
\( h(x) = d \). 
Clearly 
\( h \not \in \MANS{Q'}{\OLB{I}{}}{J} \).
As 
\( Q \preceq_J Q' \), 
this is impossible. Further
\( y \in \ELIM{Q} \), 
since we could assign to $x$ and $y$ distinct $J$-values to obtain a solution
of $Q$ in \OLA{I}{}, but this is not possible since the equation
\( x = y \) 
appears in $Q'$. Moreover 
\( y \not \in \ELIM{Q'} \). 
We assign $y$ to $x$. So elements from 
\( \ELIM{Q'} \setminus \ELIM{Q} \) 
are mapped onto elements from the set 
\( \ELIM{Q} \setminus \ELIM{Q'} \). 

The mapping so constructed is one-to-one. Consider different variables $x_1$ 
and $x_2$ from 
\( \ELIM{Q'} \setminus \ELIM{Q} \). 
If the same variable $y$ is assigned to both variables then we would have 
\( x_1 = y \) 
and 
\( x_2 = y \) 
in $Q'$. Consequently $x_1$ and $x_2$ would be bound to take the same 
$J$-values. As both do not occur in \ELIM{Q}, this is impossible. So we proved 
\( \CARD{\ELIM{Q}} \geq \CARD{\ELIM{Q'}} \). 
\END

\medskip

We can strengthen Lemma~\ref{cl:query:elim:1}. Let 
\( x \in \ELIM{Q'} \setminus \ELIM{Q} \)
and \OLA{I}{} as before. From the proof above we know that if
\( x = y \) 
is an equation in $Q'$, then there is an equation 
\( y = s \) 
in $Q$ for some $s$. The term $s$ must be the variable $x$, since otherwise
there is a solution $h$ of $Q$ in \OLA{I}{} such that 
\( h(x) \not = h(y) \).
But this is impossible since $x$ and $y$ are bound in $Q'$ to take the same
$J$-value. So if
\( \{ x_1, \ldots, x_k \} = \ELIM{Q'} \setminus \ELIM{Q} \) 
then for each equation 
\( x_i = y_i \) 
in $Q'$ we have an equation 
\( y_i = x_i \) 
in $Q$. If we put 
\[ Q'' = Q_{x_1, \ldots, x_k, y_1, \ldots, y_k} [y_1, \ldots, y_k, x_1, \ldots,
x_k], \]
then $Q''$ is isomorphic to $Q$ such that 
\( \ELIM{Q''} \supseteq \ELIM{Q'} \). 
So we proved the next lemma.

\begin{lemma}\label{cl:query:elim:2}
	Let $Q$ and $Q'$ be consistent queries in solved form. If 
	\( Q \preceq_J Q' \), 
	then there is a query $Q''$ in solved form isomorphic to $Q$ such that 
	\( \ELIM{Q''} \supseteq \ELIM{Q'} \). 
\end{lemma}

\begin{example}\label{ex:query:elim}
	Since
	\( (x = f(y)) \prec (\exists v) (x = f(v)) \),
	it is possible that 
	\( Q \prec Q' \) 
	and 
	\( \ELIM{Q} = \ELIM{Q'} \).
\end{example}

\begin{lemma}\label{cl:query:vars}
	Let $Q$ and $Q'$ be consistent queries in solved form. If 
	\( Q \preceq_J Q' \), 
	then 
	\( \VARS{Q} \supseteq \VARS{Q'} \).
\end{lemma}

{\bf Proof:}
By Lemma~\ref{cl:query:elim:2} we can suppose that 
\( \ELIM{Q} \supseteq \ELIM{Q'} \). 
Assume that there is 
\( y \in \VARS{Q'} \setminus \VARS{Q} \).
Then $y$ is a parameter of $Q'$ and hence there is an equation 
\( x = t \) 
in $Q'$, where $y$ appears in $t$. Put
\( \OLB{I}{} = \BB{L}{J}, \ldots, \BB{L}{J} \),
where
\( \CARD{\OLB{I}{}} = \CARD{Q} = \CARD{Q'} \).
Then there are solutions $h_1$ and $h_2$ of $Q$ in \OLA{I}{} which differs only
on $y$. Hence $h_1$ and $h_2$ are solutions of $Q'$ in \OLA{I}{}, as well. As
\( h_1(y) \not = h_2(y) \),
we obtain that 
\( h_1(x) \not = h_2(x) \). 
Contradiction.
\END

\begin{example}\label{ex:query:vars}
	Since
	\( (\exists u) (x = f(u,u)) \prec (\exists v_1) (\exists v_2) (x = 
		f(v_1,v_2)) \),
	it is possible that 
	\( Q \prec Q' \) 
	and 
	\( \VARS{Q} = \VARS{Q'} \).
\end{example}

Consider now two consistent queries $Q$ and $Q'$ in solved form and suppose
that
\( Q \equiv (\exists u_1) \ldots (\exists u_k) Q_0 \) 
and 
\( Q' \equiv (\exists v_1) \ldots (\exists v_l) Q'_0 \),
where 
\[ Q_0 \equiv (x_1 = s_1  \wedge \ldots x_n = s_n \wedge 
	A_1 \wedge \ldots A_m) \]
and
\[ Q'_0 \equiv (x_1 = t_1  \wedge \ldots x_n = t_n \wedge
	B_1 \wedge \ldots B_m). \] 
Put
\begin{equation}\label{eq:query:diff:1}
	\DIFF{Q}{Q'} =_{def} \bigcup_{i = 1}^{n} \DIFF{s_i}{t_i} \cup
								\bigcup_{j = 1}^{m} \DIFF{A_j}{B_j}
\end{equation}
Then the following lemma holds.

\begin{lemma}\label{cl:query:diff:1}
	If 
	\( Q \preceq_J Q' \), 
	then the set \DIFF{Q}{Q'} contains only pairs of the form
	\begin{enumerate}
		\item \( (y,y) \), 
				where $y$ is a parameter of both $Q$ and $Q'$.
		\item \( (s,v) \),
				where $v$ is a bound variable in $Q'$.
	\end{enumerate}
	Moreover the following uniqueness condition holds: for any variable $z$ there 
	is at most one term $s$ such that 
	\( (s,z) \in \DIFF{Q}{Q'} \).
\end{lemma}

\noindent
{\bf Proof:} 
Notice first that \DIFF{Q}{Q'} has the following property. Consider
\( g, h \in \VV{\LB{}}{J} \),
where
\( h = g \) 
on \VARS{Q'}. Let \OLA{I}{(g)} be a multiinterpretation 
\( g(A_1), \ldots, g(A_m) \)
based on $J$. If 
\( g \in \MANS{Q_0}{\OLB{I}{(g)}}{J} \) 
and 
\( h \in \MANS{Q_0}{\OLB{I}{(g)}}{J} \),
then for any pair 
\( (s,t) \in \DIFF{Q}{Q'} \)
we have 
\( g(s) = h(t) \).
Really, assume first that
\( (s,t) \in \DIFF{s_i}{t_i} \). 
Then 
\( g(x_i) = h(x_i) \) 
and hence 
\( g(s_i) = h(t_i) \). 
Therefore 
\( g(s) = h(t) \) 
by Lemma~\ref{cl:terms:diff}. Suppose now that
\( (s,t) \in \DIFF{A_j}{B_j} \).
Then by the form of \OLA{I}{(g)} we have
\( g(A_j) = h(B_j) \).
Consequently
\( g(s) = h(t) \).

Now let 
\( (s,t) \in \DIFF{Q}{Q'} \). 
Then $t$ must be a variable, since otherwise there is a $J$-solution $g$ of 
\( x_1 = s_1  \wedge \ldots x_n = s_n \)
such that 
\( g(s) \not \in \INST{J}{t} \).
Put
\( \OLB{I}{(g)} = g(A_1), \ldots, g(A_m) \).
Then
\( g \in \MANS{Q_0}{\OLB{I}{(g)}}{J} \) 
Since
\( Q \preceq_J Q' \), 
there is a solution $h$ of $Q'_0$ in \OLA{I}{(g)} identical to $g$ on
\VARS{Q'}. Then
\( g(s) = h(t) \). 
Contradiction. So \DIFF{Q}{Q'} may contain only pairs of the form 
\( (s,z) \), 
where $z$ is a variable.

Assume now that 
\( (s,y) \in \DIFF{Q}{Q'} \), 
where $y$ is a parameter of $Q'$. Then by Lemma~\ref{cl:query:vars} $y$ is a
parameter of $Q$, as well. We shall prove that $s$ must be the variable $y$.
Suppose that
\( s \not \equiv y \).
By similar arguments as in the previous case we find
\( g \in \VV{\LB{}}{J} \)
such that
\( g(s) \not = g(y) \)
and $Q$ is a solution of $Q_0$ in \OLA{I}{(g)}, where
\( \OLB{I}{(g)} = g(A_1), \ldots, g(A_m) \).
Therefore there is
\( h \in \MANS{Q'_0}{\OLB{I}{(g)}}{J} \)
such that
\( h = g \) 
on \VARS{Q'}. We have
\( g(s) = h(y) \). 
But this is impossible since
\( h(y) = g(y) \)
holds. 

Let $v$ be a bound variable in $Q'$. We prove that there is just one term $s$
such that 
\( (s,v) \in \DIFF{Q}{Q'} \). 
Suppose that 
\( (r_1,v) \in \DIFF{Q}{Q'} \) 
and 
\( (r_2,v) \in \DIFF{Q}{Q'} \), 
where $r_1$ and $r_2$ are distinct terms. Then we can find
\( g \in \VV{\LB{}}{J} \)
such that
\( g(r_1) \not = g(r_2) \)
and
\( g \in \MANS{Q_0}{\OLB{I}{(g)}}{J} \),
where
\( \OLB{I}{(g)} = g(A_1), \ldots, g(A_m) \).
Therefore there is
\( h \in \MANS{Q'_0}{\OLB{I}{(g)}}{J} \)
such that
\( h = g \) 
on \VARS{Q'}. Thus  
\( g(r_1) = h(v) \) 
and 
\( g(r_2) = h(v) \).
Contradiction.
\END

\medskip

As a consequence of previous lemma we have the following two claims.

\begin{lemma}\label{cl:query:diff:2}
	If 
	\( Q \approx_J Q' \) 
	in Lemma~\ref{cl:query:diff:1}, then $Q$ and $Q'$ are isomorphic.
\end{lemma}

{\bf Proof:}
Straightforward.

\begin{lemma}\label{cl:query:diff:3}
	If 
	\( Q \prec_J Q' \) 
	in Lemma~\ref{cl:query:diff:1},
	then one of the following cases should appear:
	\begin{enumerate}
		\item \( (s,v) \in \DIFF{Q}{Q'} \), 
				where $s$ is a nonvariable term and 
				\( v \in \BOUND{Q'} \).
		\item \( (y,v) \in \DIFF{Q}{Q'} \), 
				where $y$ is a parameter of $Q$  and 
				\( v \in \BOUND{Q'} \).
		\item \( (u,v_1) \in \DIFF{Q}{Q'} \) 
				and 
				\( (u,v_2) \in \DIFF{Q}{Q'} \), 
				where 
				\( u \in \BOUND{Q} \), 
				\( v_1 \in \BOUND{Q'} \), 
				\( v_2 \in \BOUND{Q'} \) 
				and 
				\( v_1 \not \equiv v_2 \).
\end{enumerate}
\end{lemma}

{\bf Proof:}
Straightforward.
\END

\medskip

The following example shows that each case in Lemma~\ref{cl:query:diff:3} may
appear.

\begin{example}\label{ex:query:diff}
	\[ (x = f(a)) \prec (\exists v) (x = f(v)) \]
	\[ (x = f(y)) \prec (\exists v) (x = f(v)) \]
	\[ (\exists u) (x = f(u,u)) \prec (\exists v_1)(\exists v_2) (x = 
		f(v_1,v_2)) \]
\end{example}

The main aim of this section is to prove the following claim.

\begin{theorem}\label{cl:query:equiv:order}
	Let $Q$ and $Q'$ be queries. If $J$ is a non-trivial model of
	\FEA{L}, then: 
	\begin{enumerate} 
		\item \( Q \approx_J Q' \) iff \( Q \approx Q' \) 
		\item \( Q \preceq_J Q' \) iff \( Q \preceq Q' \) 
	\end{enumerate}
	In particular if $Q$ and $Q'$ are equivalent queries in solved form,
	then they are isomorphic.
\end{theorem}

{\bf Proof:}
Suppose first that $Q$ and $Q'$ are queries in solved form and
\( Q \approx_J Q' \).
By a straightforward application of the previous lemmas we obtain that $Q$ and
$Q'$ are isomorphic and hence equivalent.

Assume now
\( Q \approx_J Q' \). 
Theorem~\ref{cl:Robinson} implies that there are queries $Q_1$ and $Q'_1$ in
solved form equivalent to $Q$ and $Q'$, respectively. Then
\( Q_1 \approx_J Q'_1 \) 
and hence
\( Q_1 \approx Q'_1 \). 
Thus 
\( Q \approx Q' \). 

Let
\( Q \preceq_J Q' \). 
Then 
\( (Q \wedge Q') \approx_J (Q \wedge Q) \) 
and hence by 1) 
\( (Q \wedge Q') \approx (Q \wedge Q) \).
Consequently
\( Q \preceq Q' \).
\END

\medskip

Finally we establish the following result: the preorder $\preceq$ is
well-founded i.e. there is no infinite increasing sequence
\( Q_0 \prec Q_1 \prec Q_2 \prec \ldots \)
of queries.

\begin{proposition}\label{cl:query:finite}
	Any consistent query has a finite number of generalizations, modulo 
	$\approx$.
\end{proposition}

{\bf Proof:}
Consider two consistent query $Q$ and $Q'$ in solved form and suppose
that 
\( Q \preceq Q' \), 
\( Q \equiv (\exists \overline{u}) Q_0 \) 
and 
\( Q' \equiv (\exists \overline{v}) Q'_0 \), 
where 
\begin{equation}\label{eq:query:finite:1}
	Q_0 \equiv (x_1 = s_1  \wedge \ldots x_n = s_n \wedge A_1 \wedge \ldots A_m)
\end{equation}
and
\begin{equation}\label{eq:query:finite:2}
	Q'_0 \equiv (x_1 = t_1  \wedge \ldots x_k = s_k \wedge B_1 \wedge \ldots B_m)
\end{equation}
By Lemma~\ref{cl:query:elim:1} we have
\( n \geq k \). 
Put as in (\ref{eq:query:diff:1}) 
\[ \DIFF{Q}{Q'} =_{def} \bigcup_{i = 1}^{k} \DIFF{s_i}{t_i} \cup
								\bigcup_{j = 1}^{m} \DIFF{A_j}{B_j} \]
We can prove as in Lemma~\ref{cl:query:diff:1} that \DIFF{Q_0}{Q'_0} contains
only pairs of the form
\( (r,z) \), 
where $z$ is a variable, and the {\em uniqueness condition\/} holds i.e. for any
$z$ there is at most one $r$ such that 
\( (r,z) \in \DIFF{Q_0}{Q'_0} \). 
Really, choose arbitrary but fixed non-trivial model $J$ of \FEA{L}. By
similar arguments as in the proof of 
Lemma~\ref{cl:query:diff:1} it is possible to show that (i) if 
\( (s,t) \in \DIFF{Q_0}{Q'_0} \), 
then $t$ must be a variable and (ii) if 
\( (r_1,z) \in \DIFF{Q_0}{Q'_0} \) 
and 
\( (r_2,z) \in \DIFF{Q_0}{Q'_0} \), 
then $r_1$ and $r_2$ are the same terms.

Now let $Q$ be a given consistent query and \( Q \preceq Q' \). To prove
the claim we can suppose that $Q'$ is not equivalent to \TRUE. Then there are 
queries $Q_1$ and $Q'_1$ in solved form such that (i) $Q_1$ is 
equivalent to $Q$, (ii) $Q'_1$ is equivalent to $Q'$ (iii) 
\( Q_1 \equiv (\exists \overline{u}) Q_0 \), 
where $Q_0$ is of the form (\ref{eq:query:finite:1}) 
and (iv) 
\( Q'_1 \equiv (\exists \overline{v}) Q'_0 \), 
where $Q'_0$ is of the form (\ref{eq:query:finite:2}). Then the size of terms
in $Q'_1$ is bounded by the size of corresponding terms in $Q_1$ and only the
function symbols of $Q_1$ can appear in $Q'_1$. It follows that there are only
a finite, modulo $\approx$, number of generalizations of $Q$ in solved form and
hence $Q$ has a finite, modulo $\approx$, number of generalizations.
\END

\section{Finite substitutions and existentially quantified systems of
equations}\label{sc:eqnsubst}

In this section we shall deal with finite substitutions and empty queries. 
Recall that empty queries are formulas constructed from propositional constants
and equations using the conjunction $\wedge$ and the existential quantifier
$\exists$. Actually, empty queries correspond to existentially quantified
systems of equations. From now such formulas will be called \EQ{formulas}.

We shall study two lattices: the lattice of \EQ{formulas} and
the lattice of finite substitutions. We prove that the both lattices are
isomorphic (see Theorem~\ref{cl:subst:isomorphism}). Finally, we introduce the
notion of application of substitutions to formulas and clarify its relationship
to \EQ{formulas} (see Theorem~\ref{cl:application:correspondence}).

\subsection{Lattice of \texorpdfstring{\EQ{formulas}}{ℰ-formulas}}\label{subs:eqn}

Now we restrict our attention only to \EQ{formulas}. We shall use $E$ as a
syntactical variable which varies through \EQ{formulas}. The set of
\EQ{formulas} (over \LA{}) will be denoted by \EQNA{\LB{}}.

\paragraph{}
Consider a pre-interpretation $J$. By the definition of the preorder
$\preceq_J$ ($\preceq$) and the equivalence $\approx_J$ ($\approx$) we have:
\begin{enumerate}
	\item[(a)]
			\( E \preceq_J E' \)
			iff
			\( \SOLN{J}{E} \subseteq \SOLN{J}{E'} \)
			and
			\( E \approx_J E' \)
			iff
			\( \SOLN{J}{E} = \SOLN{J}{E'} \)
	\item[(b)]
			\( E \preceq E' \)
			iff \TRUTH{\FEA{\LB{}}}{}{E \rightarrow E'}
			and
			\( E \approx E' \)
			iff \TRUTH{\FEA{\LB{}}}{}{E \leftrightarrow E'}
\end{enumerate}
As a straightforward corollary of Theorem~\ref{cl:query:equiv:order} and
Theorem~\ref{cl:Robinson} we obtain the following claims.

\begin{theorem}\label{cl:eqn:equiv:order}
	Let $E$ and $E'$ be \EQ{formulas}. If $J$ is a non-trivial model of
	\FEA{\LB{}}, then: 
	\begin{enumerate} 
		\item \( E \approx_J E' \) iff \( E \approx E' \)
		\item \( E \preceq_J E' \) iff \( E \preceq E' \) 
	\end{enumerate}
	In particular if $E$ and $E'$ are equivalent \EQ{formulas} in solved form,
	then they are isomorphic.
\end{theorem}

\begin{theorem}\label{cl:eqn:Robinson}
	The solved form algorithm applied to an \EQ{formula} will return an 
	equivalent one in solved form after finite number of steps.
\end{theorem}

\noindent
Remember that consistent \EQ{formulas} in solved form have the same set of free
variables. Thus for a given consistent \EQ{formula} $E$ there is a unique set
of variables, which is the set of free variables of its arbitrary solved form.
We shall denote this set as \KERNEL{E}.

\paragraph{}
Let \EQNB{\LB{}} be the quotient set of \EQ{formulas} under the equivalence
$\approx$. By $E_{\approx}$ we mean the class of \EQ{formulas} equivalent with
$E$. Let $\leq$ be the partial order obtained from the preorder $\preceq$.
Then:

\begin{theorem}\label{cl:eqn:lattice}
	\EQNB{\LB{}} is a complete lattice, where the smallest element is 
	\( \FALSE_{\approx} \)
	and the greatest element is 
	\( \TRUE_{\approx} \).
\end{theorem}

{\bf Proof:}
First notice that 
\( E \wedge E' \) 
is a greatest lower bound of the set 
\( \{ E, E' \} \). 
Therefore each nonempty finite set of \EQ{formulas} has a greatest lower bound.
Now let $S$ be a nonempty (possible infinite) set of \EQ{formulas}. Then from
Proposition~\ref{cl:query:finite} the set $Z$ of upper bounds of $S$ is
nonempty and finite. If
\( Z = \{ E_1, \ldots, E_n \} \),
then 
\( E_1 \wedge \ldots \wedge E_n \) 
is a lower upper bound of $S$. A greatest lower bound of $S$ we obtain as a
lower upper bound of the set of lower bounds of $S$.
\END

\paragraph{}
Now we introduce the notion of projection for \EQ{formulas}. Let 
\( E \not \equiv \FALSE \)
be an
\EQ{formula} and $X$ a finite set of variables. Let 
\( \{ z_1, \ldots, z_k \} = \VARS{E} \setminus X \),
where $z_1$, \ldots, $z_k$ occur in \VAR\ in this order. The {\em
projection\/} of $E$ onto $X$, denoted by \PROJ{E}{X}, is defined as the
\EQ{formula}
\( (\exists z_1) \ldots (\exists z_k) E \).
We put
\( \PROJ{\FALSE}{X} = \FALSE \).
By Lemma~\ref{cl:basic}
\( E \rightarrow \PROJ{E}{X} \)
is valid and hence
\( E \preceq \PROJ{E}{X} \).

\begin{lemma}\label{cl:eqn:proj:order}
	Let $E$ and $E'$ be \EQ{formulas} and $X$ a finite set of variables. Then:
	\begin{enumerate}
		\item If
				\( E \preceq E' \),
				then
				\( \PROJ{E}{X} \preceq \PROJ{E'}{X} \).
		\item If
				\( E \approx E' \),
				then
				\( \PROJ{E}{X} \approx \PROJ{E'}{X} \).
	\end{enumerate}
\end{lemma}

{\bf Proof:}
Without loss of generality we can suppose that $E$ and $E'$ are consistent
\EQ{formulas}. We first prove 2). Let
\( E \approx E' \).
We can suppose that $E'$ is in solved form. Then 
\( \VARS{E'} \subseteq \VARS{E} \).
Let
\( \{ z_1, \ldots, z_k \} = \VARS{E'} \setminus X \)
and
\( \{ z_1, \ldots, z_k, \ldots, z_n \} = \VARS{E} \setminus X \).
Then
\begin{equation}\label{eq:eqn:proj:order:1}
	\{ z_{k+1}, \ldots, z_n \} \cap \VARS{E'} = \emptyset
\end{equation}
By Lemma~\ref{cl:basic}
\( E \rightarrow (\exists z_{k+1}) \ldots (\exists z_n) E \)
is valid and since \TRUTH{\FEA{\LB{}}}{}{E' \rightarrow E} we obtain
\TRUTH{\FEA{\LB{}}}{}{E' \rightarrow (\exists z_{k+1}) \ldots (\exists z_n) E}.
Because \TRUTH{\FEA{\LB{}}}{}{E \rightarrow E'}, then by Lemma~\ref{cl:basic}
\TRUTH{\FEA{\LB{}}}{}{(\exists z_{k+1}) \ldots (\exists z_n) E \rightarrow E}
since (\ref{eq:eqn:proj:order:1}) holds. Therefore
\( E' \approx (\exists z_{k+1}) \ldots (\exists z_n) E \).
By applying Lemma~\ref{cl:basic} again we have
\[ \TRUTH{\FEA{\LB{}}}{}{(\exists z_1) \ldots (\exists z_k) E' \leftrightarrow 
	(\exists z_1) \ldots (\exists z_k) \ldots (\exists z_n) E} \]
i.e.
\( \PROJ{E'}{X} \approx \PROJ{E}{X} \).

Now assume that
\( E \preceq E' \).
By 2) we can suppose that $E$ and $E'$ are in solved form. Then 
\( \VARS{E'} \subseteq \VARS{E} \).
By similar arguments as in the proof above we can show that
\( \PROJ{E}{X} \preceq \PROJ{E'}{X} \)
holds.
\END

\begin{lemma}\label{cl:eqn:proj:answer}
	Consider arbitrary formula $F$ and a consistent \EQ{formula} $E$. Let $E'$ 
	be a projection of $E$ onto \VARS{F}. Then
	\( (\forall) (E \rightarrow F) \leftrightarrow 
	(\forall) (E' \rightarrow F) \)
	is a valid formula.
\end{lemma}

{\bf Proof:}
Let $I$ be an interpretation based on a pre-interpretation $J$. Suppose that
\TRUTH{I}{}{(\forall) (E \rightarrow F)}
or equivalently
\( \SOLN{J}{E} \subseteq \ANS{F}{J}{I} \).
Let 
\( h \in \SOLN{J}{E'} \).
Since
\( E' \equiv (\exists z_1) \ldots (\exists z_k) E \),
where
\( \{ z_1, \ldots, z_k \} = \VARS{E} \setminus \VARS{F} \),
then there is 
\( g \in \SOLN{J}{E} \)
identical to $h$ on 
\( \VAR \setminus \{ z_1, \ldots, z_k \} \).
By assumptions
\( g \in \ANS{F}{J}{I} \).
Since
\( g = h \)
on \VARS{F}, we have
\( h \in \ANS{F}{J}{I} \).
We proved that
\( (\forall) (E \rightarrow F) \rightarrow (\forall) (E' \rightarrow F) \)
is a valid formula.

Since
\( E \rightarrow E' \)
is a valid formula, then
\( (\forall) (E' \rightarrow F) \rightarrow (\forall) (E \rightarrow F) \)
is a valid formula, as well.
\END

\paragraph{}
As in \cite{LMM:unif} we establish the so-called compactness theorem. We need
to extend the definition of $\preceq$ and $\approx$ to equational formulas. Let
$F$ and $F'$ be equational formulas (over \LA{}). Then:
\begin{enumerate}
	\item[(a)]
			\( F \preceq F' \)
			iff \TRUTH{\FEA{\LB{}}}{}{F \rightarrow F'}
	\item[(b)]
			\( F \approx F' \)
			iff \TRUTH{\FEA{\LB{}}}{}{F \leftrightarrow F'}
\end{enumerate}
We shall write
\( F \prec F' \)
if
\( F \preceq F' \)
and
\( F \not \approx F' \).

\begin{theorem}[Strong Compactness]\label{cl:eqn:compactness:1}
	Let $E$ and $E_1$, \ldots, $E_n$ be \EQ{formulas}.
	\begin{enumerate}
		\item If 
				\( E_1 \prec E , \ldots, E_n \prec E \), 
				then 
				\( E_1 \vee \ldots \vee E_n \prec E \).
		\item If 
				\( E \approx E_1 \vee \ldots \vee E_n \), 
				then 
				\( E \approx E_j \) 
				for some $E_j$.
		\item If 
				\( E \preceq E_1 \vee \ldots \vee E_n \), 
				then 
				\( E \preceq E_j \) 
				for some $E_j$.
	\end{enumerate}
\end{theorem}

First we prove the following proposition. Here
\begin{enumerate}
	\item[]
			\( F \preceq_J F' \)
			iff \TRUTH{J}{}{F \rightarrow F'}
			and
			\( F \approx F' \)
			iff \TRUTH{J}{}{F \leftrightarrow F'},
\end{enumerate}
where $J$ is a pre-interpretation and $F$, $F'$ are equational formulas.

\begin{proposition}\label{cl:eqn:compactness:2}
	Suppose that \LA{}\ contains infinitely many constants and \HA\ is the 
	Herbrand pre-interpretation for $L$. Let $E$ and $E_1$, \ldots, $E_n$ be 
	\EQ{formulas} over $L$.
	\begin{enumerate}
		\item If 
				\( E_1 \prec_{\HB} E , \ldots, E_n \prec_{\HB} E \), 
				then 
				\( E_1 \vee \ldots \vee E_n \prec_{\HB} E \).
		\item If 
				\( E \approx_{\HB} E_1 \vee \ldots \vee E_n \), 
				then 
				\( E \approx_{\HB} E_j \) 
				for some $E_j$.
		\item If 
				\( E \preceq_{\HB} E_1 \vee \ldots \vee E_n \), 
				then 
				\( E \preceq_{\HB} E_j \) 
				for some $E_j$.
	\end{enumerate}
\end{proposition}

The assumption that \LA{}\ contains infinitely many constants is essential.
Consider an alphabet \LA{} consisting only from a constant $a$ and an unary
function symbol $f$. Then we have
\( (x = a) \prec_{\HB} \TRUE \)
and
\( (\exists z) (x = f(z)) \prec_{\HB} \TRUE \).
On the other side
\( ((x = a) \vee (\exists z) (x = f(z))) \approx_{\HB} \TRUE \).

\medskip

{\bf Proof of Proposition~\ref{cl:eqn:compactness:2}:} 
Let 
\( E_1 \prec_{\HB} E , \ldots, E_n \prec_{\HB} E \). 
We can suppose without loss of generality (see Section~\ref{sc:query})
that $E_1$, \ldots, $E_n$, $E$ are consistent \EQ{formulas} in solved form
such that
\( \ELIM{E_1} \supseteq \ELIM{E} \), 
\ldots, 
\( \ELIM{E_n} \supseteq \ELIM{E} \) 
and that any bound variable in $E$ does not occur freely in $E_1$, \ldots,
$E_n$. Put 
\[ X = \BOUND{E} \cup \bigcup_{i=1}^{n} (\VARS{E_i} \setminus \ELIM{E}) \]
Note that 
\( \PARAM{E} \subseteq X \), 
\( \PARAM{E_1} \subseteq X \), 
\ldots, 
\( \PARAM{E_n} \subseteq X \). Moreover
\( \ELIM{E} \cap X = \emptyset \).
Let 
\( X = \{ x_1, \ldots, x_k \} \) 
and $a_1$, \ldots, $a_k$ be distinct constants not occurring in $E_1$, \ldots, 
$E_n$. Let $E$ be of the form 
\( (\exists \overline{v}) D \). 
Then there is a $H$-solution $h$ of $D$ (and hence of $E$ as well) such
that 
\( h(x_1) = a_1 \),
\ldots, 
\( h(x_k) = a_k \). 
To prove 1) it suffices to show that $h$ is not a $H$-solution of any
$E_i$. Really, let $E_i$ be of the form
\( (\exists \overline{u}) D_i \) 
and suppose that there is 
\( g \in \SOLN{H}{D_i} \) 
such that 
\( g = h \) 
on \VARS{E_i}. We show that this is impossible. There are two cases to
consider.

Suppose that 
\( \CARD{\ELIM{E_i}} > \CARD{\ELIM{E}} \). 
Then there is an equation $x = s$ in $E_i$ such that 
\( x \not \in \ELIM{E} \). 
We have  
\( x \in X \) 
and 
\( g(x) = h(x) \). 
If $s$ is not a variable, then 
\( g(x) \not \in \INST{H}{s} \) 
and therefore 
\( g \not \in \SOLN{H}{D_i} \). 
If $s$ is a variable $z$ then 
\( z \in X \). 
Hence 
\( g \not \in \SOLN{H}{D_i} \),
since
\( x \not \equiv z \).

Assume now that 
\( \CARD{\ELIM{E_i}} = \CARD{\ELIM{E}} \). 
We can suppose that $D$ and $D_i$ are of the form 
\[ D \equiv (x_1 = t_1  \wedge \ldots x_m = t_m)\ \ {\rm and}\ \ 
	D_i \equiv (x_1 = s_1  \wedge \ldots x_m = s_m) \]
Consider \DIFF{E_i}{E} introducing in (\ref{eq:query:diff:1}). If 
\( (s,t) \in \DIFF{E_i}{E} \), 
then 
\( g(s) = h(t) \) 
(see the proof of Lemma~\ref{cl:query:diff:1}). We show that this is
impossible. According to Lemma~\ref{cl:query:diff:3} 
there are three cases to consider:
\begin{enumerate}
	\item[(i)]
			\( (r,v) \in \DIFF{E_i}{E} \), 
			where $r$ is a nonvariable term and 
			\( v \in \BOUND{E} \). 
			Then 
			\( v \in X \) 
			and hence
			\( g(r) \not = h(v) \). 
			Contradiction.
	\item[(ii)]
			\( (y,v) \in \DIFF{E_i}{E} \), 
			where 
			\( y \in \PARAM{E_i} \) 
			and 
			\( v \in \BOUND{E} \). 
			Then 
			\( v \in X \), 
			\( y \in X \) 
			and hence 
			\( g(y) \not = h(v) \). 
			Contradiction.
	\item[(iii)]
			\( (u,v_1) \in \DIFF{E_i}{E} \) 
			and 
			\( (u,v_2) \in \DIFF{E_i}{E} \), 
			where
			$v_1$ and $v_2$ are distinct bound variables in $E$.  From 
			\( g(u) = h(v_1) \) 
			and 
			\( g(u) = h(v_2) \) 
			we have 
			\( h(v_1) = h(v_2) \).
			But this is impossible since 
			\( v_1 \in X \) 
			and 
			\( v_2 \in X \).
\end{enumerate}
So we proved that if 
\( E_1 \prec_{\HB} E , \ldots, E_n \prec_{\HB} E \), 
then 
\( E_1 \vee \ldots \vee E_n \prec_{\HB} E \).

Clearly the second part of the claim immediately follows from the first. 
The third part is a trivial consequence from the second by noting that the
hypothesis implies that
\( E \approx_{\HB} (E \wedge E_1) \vee \ldots (E \wedge E_n) \).
\END

\medskip

Now we are ready to prove Theorem~\ref{cl:eqn:compactness:1}.

\medskip

{\bf Proof of Theorem~\ref{cl:eqn:compactness:1}:}
Let $F$ and $F'$ be equational formulas over \LA{}. Let \LA{c} be a canonical
language for \LA{} and \HA\ a Herbrand pre-interpretation for \LA{c}. Then
by Proposition~\ref{cl:canonical} we have:
\begin{equation}\label{eq:eqn:compactness:1}
	F \prec F'\ \ {\rm iff}\ \ F \prec_{\HB} F'
\end{equation}
Assume now that
\( E_1 \prec E , \ldots, E_n \prec E \). Then by (\ref{eq:eqn:compactness:1})
we have
\( E_1 \prec_{\HB} E , \ldots, E_n \prec_{\HB} E \). 
Proposition~\ref{cl:eqn:compactness:2} yields that  
\( E_1 \vee \ldots \vee E_n \prec_{\HB} E \). 
By applying (\ref{eq:eqn:compactness:1}) again we obtain
\( E_1 \vee \ldots \vee E_n \prec E \).
The proof of the second and the third part of the claim is similar to the proof
of Proposition~\ref{cl:eqn:compactness:2}.
\END

\subsection{Lattice of Substitutions}\label{subs:subst}

In this section we introduce the lattice of (finite) substitutions.
Theorem~\ref{cl:subst:isomorphism} states that the quotient set of
substitutions under an equivalence forms a complete lattice. Moreover, this
theorem shows us that there is a strong relationship between \EQ{formulas} and
substitutions. The crux of this relationship is the mapping between an
\EQ{formula}
\[ E \equiv (\exists z_1) \ldots (\exists z_m) (x_1 = s_1  \wedge \ldots x_n =
s_n), \] 
in solved form and a substitution
\[ \sigma = \{ \BINDA{x_1}{s_1}, \ldots, \BINDA{x_n}{s_n},
\BINDA{y_1}{y_1}, \ldots, \BINDA{y_k}{y_k} \}, \] 
where
\( y_1, \ldots, y_k \) 
are parameters of $E$. Actually this mapping gives us an isomorphism (up to an
equivalence relation) between \EQ{formulas} and substitutions. 

First we introduce the notion of $J$-instances of a substitution. Let $\sigma$
be a substitution over $X$ and $J$ a pre-interpretation. A $J$-valuation $h$ is
said to be a {\em $J$-instance} of $\sigma$ if there is
\( g \in \VV{L}{J} \) 
such that 
\( h(x) = g(\sigma(x)) \) 
for all 
\( x \in X \). 
The set of all $J$-instances of $\sigma$ is denoted by \INST{J}{\sigma}.

$J$-instances of substitutions induces a preorder $\preceq_J$ and an
equivalence $\approx_J$ as follows. Consider two substitutions $\sigma$ and
$\theta$ and let $J$ be a pre-interpretation. Then:
\begin{itemize}
	\item $\theta$ is said to be {\em more general w.r.t. $J$} than $\sigma$,
			written as 
			\( \sigma \preceq_J \theta \), 
			if 
			\( \INST{J}{\sigma} \subseteq \INST{J}{\theta} \)
	\item $\theta$ is said to be {\em equivalent\/} w.r.t. $J$ (or {\em
			$J$-equivalent\/} for short) to $\sigma$, written as 
			\( \sigma \approx_J \theta \), 
			if 
			\( \INST{J}{\sigma} = \INST{J}{\theta} \)
\end{itemize}
Note that if $\sigma$ and $\theta$ are variants, then they are $J$-equivalent.
Further if $\theta$ is an extension of $\sigma$, then
\( \INST{J}{\theta} \subseteq \INST{J}{\sigma} \).
Proposition~\ref{cl:subst:regular} gives a syntactic characteristic of
extensions of a given substitution $\sigma$ which have the same set of
$J$-instances as $\sigma$.

\begin{definition}\label{df:subst:regular}
	A substitution $\theta$ over $Y$ is a {\em regular extension\/} of a 
	substitution $\sigma$ over $X$ if (i) $\theta$ is an extension of $\sigma$ 
	(ii) $\theta$ maps the set
	\( X \setminus \DOM{\sigma} \)
	injectively into the set
	\( \VAR \setminus \RANGE{\sigma} \)
	of variables.
\end{definition}

\begin{proposition}\label{cl:subst:regular}
	Let $\theta$ be an extension of $\sigma$. If $J$ is a non-trivial model of
	\FEA{L}, then:
	\[ \theta \approx_J \sigma\ \ {\rm iff}\ \ \theta\ {\rm is\ a\ regular\ 
	extension\ of}\ \sigma \]
\end{proposition}

\noindent
{\bf Proof:}
Suppose first that $\theta$ is a regular extension of $\sigma$. Let 
\( h \in \INST{J}{\sigma} \). 
Then by the definition there is 
\( g \in \VV{L}{J} \) 
such that 
\( h(x) = g(\sigma(x)) \) 
for each
\( x \in \DOM{\sigma} \). Put
\[
	g'(y) =
		\left\{
			\begin{array}{ll}
				g(\theta^{-1}(y)) & \mbox{if \( y \in \RANGE{\theta} \setminus 
										  \RANGE{\sigma} \) } \\
				g(y)              & \mbox{otherwise}
			\end{array}
		\right.
\]
The definition of $g'$ is correct since $\theta$ is a regular extension of
$\sigma$. We have 
\( h(x) = g'(\theta(x)) \) 
for each
\( x \in \DOM{\theta} \). 
Hence 
\( \INST{J}{\sigma} \subseteq \INST{J}{\theta} \) 
and therefore 
\( \sigma \approx_J \theta \).

Assume now that $\theta$ is an extension of $\sigma$ and 
\( \INST{J}{\sigma} \subseteq \INST{J}{\theta} \). 
We shall prove that $\theta$ must be the regular extension of $\sigma$. Note
first that if for some
\( g, g' \in \VV{L}{J} \) is 
\( h(x) = g(\sigma(x)) \) 
for all 
\( x \in \DOM{\sigma} \) 
and 
\( h(x) = g'(\theta(x)) \) 
for all 
\( x \in \DOM{\theta} \), 
then 
\( g'(y) = g(y) \) 
for all 
\( y \in \RANGE{\sigma} \). 
Now suppose that 
\( x \in X \), 
where 
\( X = \DOM{\theta} \setminus \DOM{\sigma} \). 
Then 
\( \theta(x) \) 
must be a variable, since otherwise there is a $J$-instance $h$ of $\sigma$
such that
\( h(x) \not \in \INST{J}{\theta(x)} \). 
Clearly 
\( h \not \in \INST{J}{\theta} \). 
Contradiction. Further, since there is 
\( h \in \INST{J}{\sigma} \) 
such that 
\( h(x_1) \not = h(x_2) \) 
for any distinct variables $x_1$ and $x_2$ from $X$, $\theta$ is one-to-one on 
the set $X$. Now suppose that there is 
\( x \in X \) 
and 
\( x' \in \DOM{\sigma} \) 
such that 
\( \sigma(x') \) 
and 
\( \theta(x) \) 
are the same variable $y$. 
Consider arbitrary $J$-valuation $g$. Then there is $h$ such that 
\( h(z) = g(\sigma(z)) \) 
for all 
\( z \in \DOM{\sigma} \) 
and 
\( h(x) \not = g(y) \). 
By assumptions for some $g'$, where 
\( h(z) = g'(\theta(z)) \) 
for all 
\( z \in \DOM{\theta} \), 
we have 
\( h(x) = g'(\theta(x)) = g'(y) = g(y) \). 
Contradiction. Finally suppose that there are variables
\( x \in X \) 
and 
\( x' \in \DOM{\sigma} \) 
such that $\theta(x)$ is a variable $y$ which does not
occur in a nonvariable term 
\( t \equiv \sigma(x') \). 
Consider arbitrary
\( g \in \VV{L}{J} \). 
Then there is $h$ such that 
\( h(z) = g(\sigma(z)) \) 
for all 
\( z \in \DOM{\sigma} \) 
and 
\( h(x) = g(t) \). 
By assumptions for some $g'$, where 
\( h(z) = g'(\theta(z)) \) 
for all 
\( z \in \DOM{\theta} \), 
we have 
\( g(y) = g'(y) = g'(\theta(x)) = h(x) = g(t) \). 
Therefore the equation 
\( y = t \) 
is $J$-solvable. Contradiction. So we proved that $\theta$ is a regular
extension of $\sigma$.
\END

\medskip

Now we introduce standard preorder $\preceq$ and equivalence $\approx$ using a
composition of substitutions:
\begin{itemize}
	\item $\theta$ is {\em more general} than $\sigma$, written as 
			\( \sigma \preceq \theta \), 
			if there is $\tau$ such that 
			\( \sigma' = \theta' \tau \), 
			where $\sigma'$ and $\theta'$ are regular extensions of $\sigma$ and 
			$\theta$ respectively over the same domain 
	\item $\theta$ is {\em equivalent} to $\sigma$, written as 
			\( \theta \approx \sigma \), 
			if 
			\( \theta \preceq \sigma \) 
			and 
			\( \sigma \preceq \theta \).
\end{itemize}
Clearly if $\sigma$ and $\theta$ are variants, then they are equivalent. The
following theorem shows that as in the case of terms and \EQ{formulas}
preorders $\preceq_J$ and $\preceq$ ($\approx_J$ and $\approx$ as well) are
identical, provided that $J$ is a non-trivial model of \FEA{L}.

\begin{theorem}\label{cl:subst:preorder}
	Let $\sigma$ and $\theta$ be substitutions. If $J$ is a non-trivial model of
	\FEA{L}, then:
	\begin{enumerate}
		\item \( \sigma \preceq_J \theta \) 
				iff 
				\( \sigma \preceq \theta \) 
		\item \( \sigma \approx_J \theta \) 
				iff 
				\( \sigma \approx \theta \) 
	\end{enumerate}
\end{theorem}

\noindent
{\bf Proof:}
Let \( \sigma \preceq_J \theta \). Due to Proposition~\ref{cl:subst:regular} we
can suppose that $\sigma$ and $\theta$ are substitutions over the same domain 
\( X = \{ x_1, \ldots, x_n \} \). 
Put 
\[ \DIFF{\sigma}{\theta} =_{def} \bigcup_{i = 1}^{n} \DIFF{s_i}{t_i}, \]
where 
\( s_i \equiv \sigma(x_i) \) 
and 
\( t_i \equiv \theta(x_i) \).

Let 
\( (s,t) \in \DIFF{\sigma}{\theta} \). 
Then $t$ must be a variable since otherwise there is 
\( g \in \VV{L}{J} \) 
such that 
\( g(s) \not \in \INST{J}{t} \). 
If 
\( h(x) = g(\sigma(x)) \) 
for 
\( x \in X \) 
then 
\( h \in \INST{J}{\sigma} \)
and 
\( h \not \in \INST{J}{\theta} \). 
Contradiction.

Let 
\( (s,y) \in \DIFF{\sigma}{\theta} \) 
and 
\( (s',y) \in \DIFF{\sigma}{\theta} \). 
Then $s$ and $s'$ must be the same terms since otherwise there is 
\( g \in \VV{L}{J} \) 
such that 
\( g(s') \not = g(s) \). 
If 
\( h(x) = g(\sigma(x)) \) 
for 
\( x \in X \) 
then 
\( h \in \INST{J}{\sigma} \)
and 
\( h \not \in \INST{J}{\theta} \). 
Contradiction. 

Now let $\tau$ be a substitution over \RANGE{\theta} such that for any 
\( y \in \RANGE{\theta} \) 
it holds 
\( \tau(y) \equiv s \), 
where 
\( (r,y) \in \DIFF{\sigma}{\theta} \). 
Clearly $\tau$ is well defined. Then we have 
\( \sigma = \theta \circ \tau \) 
and hence 
\( \sigma \preceq \theta \). 
\END

\medskip

As a straightforward corollary of the proof above we have the following claim.

\begin{corollary}\label{cl:subst:variant}
	If in the previous theorem $\sigma$ and $\theta$ are over same domain, then:
	\[ \sigma \approx_J \theta\ \ {\rm iff}\ \ \sigma \approx \theta\ \ {\rm 
	iff}\ \ \sigma\ {\rm and}\ \theta\ {\rm are\ variants} \]
\end{corollary}

The following claim states that the projection preserves the preorder $\preceq$
and the equivalence $\approx$.

\begin{theorem}\label{cl:subst:restriction}
	Consider substitutions $\sigma$ and $\theta$. Then:
	\begin{enumerate}
		\item If 
				\( \sigma \preceq \theta \),
				then
				\( \RESTR{\sigma}{X} \preceq \RESTR{\theta}{X} \).
		\item If 
				\( \sigma \approx \theta \),
				then
				\( \RESTR{\sigma}{X} \approx \RESTR{\theta}{X} \).
	\end{enumerate}
\end{theorem}

{\bf Proof:}
Obviously if $\theta$ is a regular extension of $\sigma$ then they are
equivalent. Now let
\( \sigma \preceq \theta \).
Then there is $\tau$ such that 
\( \sigma' = \theta' \tau \), 
where $\sigma'$ and $\theta'$ are regular extensions of $\sigma$ and $\theta$
respectively over the same domain. We have
\( \RESTR{\sigma'}{X} = \RESTR{\theta'}{X} \circ \RESTR{\tau}{Y} \),
where
\( Y = \RANGE{\RESTR{\theta'}{X}} \),
and hence
\( \RESTR{\sigma'}{X} \preceq \RESTR{\theta}{X'} \).
For \RESTR{\sigma'}{X} and \RESTR{\theta'}{X} are regular extensions of
\RESTR{\sigma}{X} and \RESTR{\theta}{X} respectively, we obtain that
\( \RESTR{\sigma}{X} \preceq \RESTR{\theta}{X} \)
holds.
\END
\medskip

Clearly equivalent substitutions may have different domains.  Nevertheless we 
can find for any substitutions $\sigma$ a minimal set $X$ of variables (under
set inclusion), denoted \KERNEL{\sigma}, having the following property: the
restriction of $\sigma$ onto $X$ is equivalent to $\sigma$. Really, consider a
a set
\[ \KK{\sigma} = \{ X \subseteq \VAR \mid \RESTR{\sigma}{X} \approx \sigma \}.
\]
Then the existence of such set is a straightforward consequence of the
following properties of \KK{\sigma}:
\begin{itemize}
	\item \KK{\sigma} is nonempty, since 
			\( \DOM{\sigma} \in \KK{\sigma} \)
	\item if 
			\( X \in \KK{\sigma} \),
			then $\sigma$ is a regular extension of \RESTR{\sigma}{X}
	\item \KK{\sigma} has a finite intersection property i.e. if 
			\( X_i \in \KK{\sigma} \), 
			then 
			\( X_1 \cap X_2 \in \KK{\sigma} \).
\end{itemize}
The last follows by noting that $\sigma$ is a regular extension of 
\RESTR{\sigma}{(X_1 \cap X_2)}. Now we put
\[
	\KERNEL{\sigma} = \bigcap \{ X | X \in \KK{\sigma} \}
\]
The following theorem establishes properties of kernels.

\begin{theorem}\label{cl:subst:kernel}
	Consider substitutions $\sigma$ and $\theta$. Then:
	\begin{enumerate}
	\item	$\sigma$ is a regular extension of \RESTR{\sigma}{\KERNEL{\sigma}}.
	\item	If
			\( \sigma \preceq \theta \),
			then
			\( \KERNEL{\sigma} \supseteq \KERNEL{\theta} \).
	\item	If
			\( \sigma \approx \theta \),
			then
			\( \KERNEL{\sigma} = \KERNEL{\theta} \).
\end{enumerate}
\end{theorem}

{\bf Proof:}
Let
\( \sigma \preceq \theta \).
To prove 2) it sufficient to show that
\( \KK{\sigma} \subseteq \KK{\theta} \)
or equivalently for every $X$ if $\sigma$ is a regular extension of
\RESTR{\sigma}{X}, then $\theta$ is a regular extension of \RESTR{\theta}{X} as
well. Now let $\sigma$ is a regular extension of
\RESTR{\sigma}{X}. By assumptions there is $\tau$ such that
\( \sigma' = \theta' \tau \), 
where $\sigma'$ and $\theta'$ are regular extensions of $\sigma$ and $\theta$
respectively over the same domain. Then $\sigma'$ is a regular extension of
\RESTR{\sigma}{X} and hence $\theta'$ is a regular extension of
\RESTR{\theta'}{(X \cap \DOM{\theta})}. But
\( \RESTR{\theta'}{(X \cap \DOM{\theta})} \equiv \RESTR{\theta}{X} \)
and therefore $\theta$ is a regular extension of \RESTR{\theta}{X}.
\END
\medskip

Let \SUBFALSE\ be an arbitrary object that is not element of \SUBA{L}. Let
\SUBB{L} be the set 
\( \SUBA{L} \cup \{ \SUBFALSE \} \).
We extend the partial ordering $\preceq$ and the equivalence $\approx$ to
\SUBB{L}\ by requiring \SUBFALSE\ to be the smallest element of \SUBB{L}. We 
denote by \SUBC{L}\ the new quotient set and by $\sigma_{\approx}$ the
equivalence class in which $\sigma$ lies. We put 
\( \INST{J}{\SUBFALSE} = \emptyset \).

\begin{theorem}\label{cl:subst:isomorphism}
	The lattice \SUBC{L}\ is isomorphic to the lattice \EQNB{L}. In 
	particular, \SUBC{L}\ is a complete lattice with \SUBEMPTYB\ as the 
	greatest element and with \SUBFALSE\ as the smallest element.
\end{theorem}

\noindent
{\bf Proof:}
The isomorphic mapping \IM\ between \EQNB{L} and \SUBC{L} is defined as
follows:
\begin{enumerate}
	\item \( \IM(True_{\approx}) = \SUBEMPTYB \) 
			and 
			\( \IM(False_{\approx}) = \SUBFALSE \)
	\item \( \IM(E_{\approx}) = \sigma_{\approx} \), 
			if 
			\begin{equation}\label{eq:isomorphism:1}
				E \equiv (\exists z_1) \ldots (\exists z_m) (x_1 = s_1 \wedge 
				\ldots x_n = s_n)
			\end{equation}
			is in solved form and
			\begin{equation}\label{eq:isomorphism:2}
				\sigma = \{ \BINDA{x_1}{s_1}, \ldots, \BINDA{x_n}{s_n},
				\BINDA{y_1}{y_1}, \ldots, \BINDA{y_k}{y_k} \},
			\end{equation}
			where
			\( y_1, \ldots, y_k \) 
			are parameters of $E$.
\end{enumerate}
To prove that \IM\ is an isomorphism between \EQNB{L} and \SUBC{L} it is
sufficient to show that (i) if $J$ is a model of \FEA{L}, then 
\( \SOLN{J}{E} = \INST{J}{\sigma} \), 
where $E$ and $\sigma$ are from 2) and (ii) for any $\sigma$ there is $E$ such
that
\( \IM(E_{\approx}) = \sigma_{\approx} \). 

(i) Suppose 
\( h \in \SOLN{J}{E} \). 
Then there is $g$ identical to $h$ on \VARS{E} such that 
\( g(x_i) = g(s_i) \) 
for 
\( i = 1, \ldots, n \). 
Then 
\( h(x) = g(\sigma(x)) \) 
for 
\( x \in \DOM{\sigma} \). 
Thus we have 
\( h \in \INST{J}{\sigma} \). 
Assume now that 
\( h \in \INST{J}{\sigma} \). 
Then there is $g$ such that 
\( h(x_i) = g(s_i) \) 
and 
\( h(y_j) = g(y_j) \) 
for any $x_i$ and $y_j$. Put 
\( g'(u) = h(u) \) 
on \VARS{E} and 
\( g'(u) = g(u) \) 
otherwise. Then 
\( g'(y_j) = g(y_j) \) 
and hence 
\( g'(s_i) = g(s_i) \) 
for any $s_i$. We have 
\( g'(x_i) = h(x_i) = g(s_i) = g'(s_i) \) 
and therefore 
\( h \in \SOLN{J}{E} \).

(ii) Consider arbitrary substitution 
\[ \sigma = \{ \BINDA{x_1}{s_1}, \ldots, \BINDA{x_n}{s_n} \}. \]
We can suppose due to Corollary~\ref{cl:subst:variant} that \DOM{\sigma} and
\RANGE{\sigma} are disjoint sets. Put
\[ E \equiv (\exists z_1) \ldots (\exists z_m) (x_1 = s_1 \wedge \ldots x_n = 
s_n), \] 
where 
\( \{ z_1, \ldots, z_m \} = \RANGE{\sigma} \). 
By similar arguments as in (i) we obtain that 
\( \SOLN{J}{E} = \INST{J}{\sigma} \), 
where $J$ is arbitrary model \FEA{L}, and hence 
\( \IM(E_{\approx}) = \sigma_{\approx} \). 
\END

\medskip

As a straightforward corollary from the proof above we obtain:

\begin{corollary}\label{cl:subst:isomorhism:1}
	Suppose that 
	\( \IM(E_{\approx}) = \sigma_{\approx} \). 
	Then:
	\begin{enumerate}
		\item \( \SOLN{J}{E} = \INST{J}{\sigma} \) 
				for any model $J$ of \FEA{L}.
		\item \( \KERNEL{E} = \KERNEL{\sigma} \), 
				provided that $E$ is consistent.
		\item \( \IM((\RESTR{E}{X})_{\approx}) = (\RESTR{\sigma}{X})_{\approx} \).
	\end{enumerate}
\end{corollary}

\subsection{Application of Substitutions to Formulas}\label{subs:application}

In Section~\ref{sc:basic} we introduced the notion of application of
substitutions to formulas. Recall that if
\( \sigma = \{ \BINDA{x_1}{s_n}, \ldots, \BINDA{x_n}{s_n} \} \)
then
\( F\sigma \)
is defined as
\( F_{x_1, \ldots, x_n}[s_1, \ldots, s_n] \),
provided that $\sigma$ is applicable to $F$. Now we generalize this notion for
arbitrary substitutions.

Consider a formula $F$ and a substitution $\sigma$. Suppose first that
\( \DOM{\sigma} = \VARS{F} \).
The application is defined by structural induction as follows:
\begin{enumerate}
	\item Let $F$ is an atom
			\( p(s_1, \ldots, s_n) \).
			Then
			\( F \sigma =_{def}  p(s_1 \sigma, \ldots, s_n \sigma) \).
	\item Let $F$ is a formula of the form
			\( \neg G \).
			Then
			\( F \sigma =_{def}  \neg (G \sigma) \).
	\item Let $F$ is a formula of the form
			\( G\ b\ H \),
			where $b$ is $\wedge$, $\vee$ or $\rightarrow$.
			Then
			\( F \sigma =_{def}  G \sigma_1\ b\ H \sigma_2 \),
			where $\sigma_1$ or $\sigma_2$ is a restriction of $\sigma$ onto
			\VARS{G} or \VARS{H}, respectively.
	\item Let $F$ is a formula of the form
			\( (\exists x) G \).
			\begin{enumerate}
				\item Let 
						\( x \not \in \RANGE{\sigma} \).
						We put
						\( F \sigma =_{def} (\exists x) (G \sigma') \),
						where 
						\[
							\sigma' = \left\{ \begin{array}{ll}
														\sigma \cup \BINDB{x}{x} & \mbox{if
															\( x \in \VARS{G} \)} \\
														\sigma                   & \mbox{otherwise}
													\end{array}
										 \right.
						\]
				\item	Let
						\( x \in \RANGE{\sigma} \).
						Let
						\( (\exists y) G_{x}[y] \)
						be a variant of $F$, where $y$ is the first variable in 
						\VAR\ such that $y$ does not occur in \RANGE{\sigma}.
						We put
						\( F \sigma =_{def} (\exists y) (G_{x}[y] \sigma') \),
						where
						\[
							\sigma' = \left\{ \begin{array}{ll}
														\sigma \cup \BINDB{y}{y} & \mbox{if
															\( x \in \VARS{G} \)} \\
														\sigma                   & \mbox{otherwise}
													\end{array}
										 \right.
						\]
			\end{enumerate}
	\item The case when $F$ is a formula of the form
			\( (\forall x) G \)
			is similar to previous one.
\end{enumerate}

\begin{example}\label{ex:application:1}
	Let
	\( \VAR = \{x, y, z, u, \ldots \} \).
	Consider a  formula
	\( F = (\exists z) p(x,y,z) \)
	and a substitution
	\( \sigma = \{ \BINDA{x}{z}, \BINDA{y}{x} \} \).
	Then
	\( F \sigma = (\exists u) p(z,x,u) \).
\end{example}

Now consider an arbitrary substitution $\sigma$. First we restrict $\sigma$ to
the set \VARS{F} to obtain a substitution $\sigma'$ over
\( \DOM{\sigma} \cap \VARS{F} \).
Now let $\sigma''$ be a regular extension of $\sigma'$ onto \VARS{F}. We put
\( F \sigma =_{def} F \sigma'' \),
where the right hand side is correctly defined since
\( \DOM{\sigma''} = \VARS{F} \).
Clearly the application of $\sigma$ to $F$ depends on the choice of $\sigma''$.
Nevertheless we can select $\sigma''$ uniquely by applying the following
procedure:
{\em \begin{tabbing}
===\==\=\kill
\>	\( \sigma'' := \sigma' \) \\
\>	\underline{\bf while}
		\( \VARS{F} \setminus \DOM{\sigma''} \) 
		is nonempty \underline{\bf do} \\
\>\>	\underline{\bf let}
	 		$x$ be the first variable occurring in 
			\( \VARS{F} \setminus \DOM{\sigma''} \) \\
\>\>	\underline{\bf let}
			\(
				y\ \mbox{be}
					\left\{ \begin{array}{ll}
									x & \mbox{if \( x \not \in \RANGE{\sigma''} \)} \\
									\mbox{the first variable not occurring in \RANGE{\sigma''}}
									  & \mbox{otherwise}
							  \end{array}
					\right.
			\) \\
\>\>	\( \sigma'' := \sigma'' \cup \BINDB{x}{y} \) \\
\>	\underline{\bf enddo}
\end{tabbing}}
\noindent
Obviously $\sigma''$ is a regular extension of $\sigma'$.

\begin{example}\label{ex:application:2}
	Let
	\( \VAR = \{x, y, z, u, \ldots \} \).
	\begin{enumerate}
		\item Suppose that
				\( F = (\exists z) p(x,y,z) \)
				and
				\( \sigma = \{ \BINDA{x}{f(z,x)}, \BINDA{z}{g(x,y,z)} \} \).
				Then
				\( \sigma' = \{ \BINDA{x}{f(z,x)} \} \)
				and
				\( \sigma'' = \{ \BINDA{x}{f(z,x)}, \BINDA{y}{y} \} \).
				We have
				\( F \sigma = (\exists u) p(f(z,x),y,u) \).
		\item Suppose that
				\( F = (\exists z) p(x,y,z) \)
				and
				\( \sigma = \{ \BINDA{x}{f(z,y)}, \BINDA{z}{g(x,y,z)} \} \).
				Then
				\( \sigma' = \{ \BINDA{x}{f(z,y)} \} \)
				and
				\( \sigma'' = \{ \BINDA{x}{f(z,y)}, \BINDA{y}{x} \} \).
				We have
				\( F \sigma = (\exists u) p(f(z,y),x,u) \).
	\end{enumerate}
\end{example}

\paragraph{}
The main aim of this section is to prove
Theorem~\ref{cl:application:application}. To prove it we first need the
following lemmas.

\begin{lemma}\label{cl:application:1}
	Consider formulas $F$ and $F'$ and a substitution $\sigma$.
	\begin{enumerate}
		\item Let
				\( \sigma = \{ \BINDA{x_1}{s_n}, \ldots, \BINDA{x_n}{s_n} \} \)
				be applicable to $F$. Then
				\[ F \sigma = F_{x_1, \ldots, x_n}[s_1, \ldots, s_n]. \]
		\item If $F$ and $F'$ are variants, then 
				\( F \sigma \)
				and
				\( F' \sigma \)
				are variants, as well.
	\end{enumerate}
\end{lemma}

{\bf Proof:}
(1) Straightforward. (2) If $F$ and $F'$ are variants, then
\( \VARS{F} = \VARS{F'} \)
and thus we can suppose that
\( \DOM{\sigma} = \VARS{F} \).
The claim is obtained by structural induction on formulas. The proof is long
and straightforward. We shall omit them.
\END

\begin{lemma}\label{cl:application:2}
	Let $I$ be an interpretation over $J$. Consider a formula $F$ and 
	a substitution $\sigma$. Suppose that
	\( \sigma = \{ \BINDA{x_1}{s_n}, \ldots, \BINDA{x_n}{s_n} \} \)
	is applicable to $F$ and 
	\( \DOM{\sigma} = \VARS{F} \).
	Let
	\( g,h \in \VV{L}{J} \).
	If 
	\( h(x) = g(\sigma(x)) \) 
	for any 
	\( x \in \DOM{\sigma} \), 
	then:
	\[ 
		\TRUTH{I}{h}{F}\ {\rm iff}\
		\TRUTH{I}{g}{F_{x_1, \ldots, x_n}[s_1, \ldots, s_n]}
	\]
\end{lemma}

71
{\bf Proof:}
We prove this claim by the structural induction on formulas. (i) Let $F$ is an
equation
\( r = t \). 
Then we obtain that 
\TRUTH{I}{h}{r = t} 
iff 
\( h(r) = h(t) \)  
iff 
\( g(r \sigma) = g(t \sigma) \)
iff 
\TRUTH{I}{g}{(r = t)_{x_1, \ldots, x_n}[s_1, \ldots, s_n]}.
(ii) Let $F$ is an atom 
\( p(s_1, \ldots, s_m) \). 
Then we have 
\TRUTH{I}{h}{p(s_1, \ldots, s_m)} 
iff 
\( p(h(s_1), \ldots, h(s_m)) \in I \) 
iff 
\( p(g(s_1 \sigma), \ldots, g(s_m \sigma)) \in I \) 
iff 
\TRUTH{I}{g}{p(s_1, \ldots, s_m)_{x_1, \ldots, x_n}[s_1, \ldots, s_n]}, 
since 
\( h(s_i) = g(s_i \sigma) \).
(iii) Let 
\( F \equiv \neg G \). 
Then
\TRUTH{I}{h}{\neg G} 
iff 
\NOTTRUTH{I}{h}{G} 
iff 
\NOTTRUTH{I}{g}{G_{x_1, \ldots, x_n}[s_1, \ldots, s_n]} 
iff
\TRUTH{I}{g}{F_{x_1, \ldots, x_n}[s_1, \ldots, s_n]}. 
(iv) Let $F$ is a conjunction of $G$ and $H$. Denote by $\sigma'$ or $\sigma''$
the restriction of $\sigma$ to \VARS{G} or \VARS{H}, respectively. Let
\[ 
	\sigma' = \{ \BINDA{y_1}{s'_1}, \ldots, \BINDA{y_k}{s'_k} \}\ 
	{\rm and}\ 
	\sigma'' = \{ \BINDA{z_1}{s''_n}, \ldots, \BINDA{z_l}{s''_l} \}.
\]
Then using induction hypothesis we have
\TRUTH{I}{h}{G \wedge H} 
iff 
\TRUTH{I}{h}{G} and \TRUTH{I}{h}{H} 
iff 
\TRUTH{I}{g}{G_{y_1, \ldots, y_k}[s'_1, \ldots, s'_k]} 
and 
\TRUTH{I}{g}{H_{z_1, \ldots, z_l}[s''_1, \ldots, s''_l]} 
iff 
\TRUTH{I}{g}{F_{x_1, \ldots, x_n}[s_1, \ldots, s_n]}.
(v) By similar arguments we can prove the claim if $F$ is of the form 
\( G \vee H \), 
\( G \rightarrow H \) 
or 
\( G \leftrightarrow H \).
(vi) Let 
\( F \equiv (\exists y) G \). 
Then \TRUTH{I}{h}{(\exists y) G} iff 
\TRUTH{I}{\CCB{h}{y}{d}}{G}
for some 
\( d \in \UU{L}{J} \).
Let
\[
	\tau = \left\{ \begin{array}{ll}
						\sigma \cup \BINDB{y}{y} & \mbox{if \( y \in \VARS{G} \)} \\
						\sigma                   & \mbox{otherwise}
							\end{array}
			 \right.
\]
Then
\( \CCA{h}{y}{d}(x) = \CCA{g}{y}{d}(\tau(x)) \) 
for any 
\( x \in \DOM{\tau} \). 
We have by induction hypothesis
\[
	\TRUTH{I}{\CCB{h}{y}{d}}{G}\ \mbox{iff}
			\left\{ \begin{array}{ll}
							\TRUTH{I}{\CCB{g}{y}{d}}{G_{x_1, \ldots, x_n, y}[s_1, \ldots, s_n, y]}
						 		& \mbox{if \( y \in \VARS{G} \)} \\
							\TRUTH{I}{\CCB{g}{y}{d}}{G_{x_1, \ldots, x_n}[s_1, \ldots, s_n]}
								& \mbox{otherwise}
					  \end{array}
			\right.
\]
Hence 
\TRUTH{I}{h}{(\exists y) G} 
iff 
\TRUTH{I}{g}{F_{x_1, \ldots, x_n}[s_1, \ldots, s_n]}.
(vii) Using the same arguments we can prove that the claim holds when 
\( F \equiv (\forall y) G \).
\END

\begin{lemma}\label{cl:application:3}
	Let $I$ be an interpretation over $J$. Consider a formula $F$ and 
	a substitution $\sigma$. Let $\sigma'$ be a restriction of $\sigma$ onto 
	\VARS{F}. Then:
	\[
		\INST{J}{\sigma} \subseteq \ANS{F}{I}{J}\ {\rm iff}\
		\INST{J}{\sigma'} \subseteq \ANS{F}{I}{J}
	\]
\end{lemma}

{\bf Proof:}
Since
\( \sigma \preceq \sigma' \),
\( \INST{J}{\sigma'} \subseteq \ANS{F}{I}{J} \)
implies
\( \INST{J}{\sigma} \subseteq \ANS{F}{I}{J} \).
Let
\( \INST{J}{\sigma} \subseteq \ANS{F}{I}{J} \)
and
\( h \in \INST{J}{\sigma'} \).
Then there is
\( g \in \INST{J}{\sigma} \)
such that
\( g = h \)
on \VARS{F}. By assumption
\( g \in \ANS{F}{I}{J} \)
and hence
\( h \in \ANS{F}{I}{J} \).
\END

\begin{lemma}\label{cl:application:4}
	Let $I$ be an interpretation over $J$. Consider a formula $F$ and a 
	substitution $\sigma$. Then:
	\[
		\TRUTH{I}{}{F \sigma}\ {\rm iff}\ 
		\INST{J}{\sigma} \subseteq \ANS{F}{I}{J}
	\]
\end{lemma}

{\bf Proof:}
By the definition of application and Lemma~\ref{cl:application:3} we can
suppose without lost of generality that
\( \DOM{\sigma} = \VARS{F} \).
Let
\( \sigma = \{ \BINDA{x_1}{s_n}, \ldots, \BINDA{x_n}{s_n} \} \).
Using Lemma~\ref{cl:application:1}(2) we can assume $\sigma$ is
applicable to $F$. Consequently by Lemma~\ref{cl:application:1}(1)
\( F \sigma = F_{x_1, \ldots, x_n}[s_1, \ldots, s_n] \).
So we must prove that
\[
	\TRUTH{I}{}{F_{x_1, \ldots, x_n}[s_1, \ldots, s_n]}\ {\rm iff}\ 
	\INST{J}{\sigma} \subseteq \ANS{I}{J}{F}
\]
Suppose that
\TRUTH{I}{}{F_{x_1, \ldots, x_n}[s_1, \ldots, s_n]}. 
Let 
\( h \in \INST{J}{\sigma} \). 
Then there is $g$ such that 
\( h(x) = g(\sigma(x)) \) 
for 
\( x \in \DOM{\sigma} \). 
By Lemma~\ref{cl:application:2}
\TRUTH{I}{g}{F_{x_1, \ldots, x_n}[s_1, \ldots, s_n]}  
and hence \TRUTH{I}{h}{F}. Assume now that 
\( \INST{J}{\sigma} \subseteq \ANS{I}{J}{F} \). 
Consider arbitrary 
\( g \in \VV{L}{J} \). 
Then there is 
\( h \in \INST{J}{\sigma} \) 
such that 
\( h(x) = g(\sigma(x)) \) 
for 
\( x \in \DOM{\sigma} \). 
By Lemma~\ref{cl:application:2} again we have \TRUTH{I}{h}{F} and hence 
\TRUTH{I}{g}{F_{x_1, \ldots, x_n}[s_1, \ldots, s_n]}.
\END

\paragraph{}
We are now in position to prove the desired theorem.

\begin{theorem}\label{cl:application:application}
	The application of substitutions to formulas has the following properties:
	\begin{enumerate}
		\item[(a)]
				If
				\( \sigma \preceq \theta \),
				then
				\( (\forall) F \sigma \rightarrow (\forall) F \theta \)
				is valid.
		\item[(b)]
				If
				\( \sigma \approx \theta \),
				then
				\( (\forall) F \sigma \leftrightarrow (\forall) F \theta \)
				is valid.
		\item[(c)]
				If $\sigma'$ is a restriction of $\sigma$ onto \VARS{F}, then
				\( (\forall) F \sigma \leftrightarrow (\forall) F \sigma' \)
				is valid.
	\end{enumerate}
\end{theorem}

{\bf Proof:}
Straightforward by applying the previous lemmas.
\END

\paragraph{}
We conclude this section by showing that there is a interesting relationship
between applications of substitutions and \EQ{formulas}.

\begin{theorem}\label{cl:application:correspondence}
	Let $F$ be a formula. Suppose that a substitution $\sigma$ corresponds to 
	an \EQ{formula} $E$ in the isomorphic mapping between lattices \EQNB{L} and 
	\SUBC{L}. Then
	\[ \TRUTH{\FEA{L}}{}{F \sigma}\ {\rm iff}\ 
	\TRUTH{\FEA{L}}{}{E \rightarrow F}, \]
	or equivalently
	\[ \TRUTH{\FEA{L}}{}{(\forall) F \sigma \leftrightarrow (\forall) (E 
	\rightarrow F)}. \]
\end{theorem}

{\bf Proof:}
Let $I$ over $J$ be a model of \FEA{L}. Then by Lemma~\ref{cl:application:3}
and Corollary~\ref{cl:subst:isomorhism:1} we have
\TRUTH{I}{}{F \sigma}
iff 
\( \INST{J}{\sigma} \subseteq \ANS{I}{J}{F} \)
iff
\( \SOLN{J}{E} \subseteq \ANS{I}{J}{F} \)
iff
\TRUTH{I}{}{E \rightarrow F}.
\END

\end{document}